\documentclass[11pt,a4paper]{article}
\usepackage[text={16.5cm,24cm},centering]{geometry}

\usepackage{amsmath}
\usepackage{amsthm}
\usepackage[utf8]{inputenc}
\usepackage{tgtermes,tgheros,tgcursor}
\usepackage[varg]{newtxmath}
\usepackage{hyperref}
\hypersetup{%
 colorlinks=true,%
 linkcolor=blue,
 citecolor=blue,
}
\usepackage{graphicx}
\usepackage{graphbox}
\graphicspath{{figure/}}
\usepackage[%
  format=plain,
  margin=1cm,
  font=small
]{caption}
\usepackage{here}
\usepackage{physics}
\usepackage{xspace}
\numberwithin{equation}{section}

\newcommand{\Zb}{\mathbb{Z}}
\newcommand{\Rb}{\mathbb{R}}
\newcommand{\confA}{\textsf{A}\xspace}
\newcommand{\confB}{\textsf{B}\xspace}
\newcommand{\confC}{\textsf{C}\xspace}

\begin{document}
\begin{center}
  \begin{flushright}
    OU-HET 1105
  \end{flushright}
  \vspace{8ex}
  {\LARGE\bfseries \boldmath Non-invertible topological defects in 4-dimensional $\Zb_2$ pure lattice gauge theory}\\
  \vspace{4ex}
  {\Large Masataka Koide, Yuta Nagoya, and Satoshi Yamaguchi}\\
  \vspace{2ex}
  {\itshape Department of Physics, Graduate School of Science, 
  \\
  Osaka University, Toyonaka, Osaka 560-0043, Japan}\\
  \vspace{1ex}
  {\footnotesize Email: \texttt{mkoide@het.phys.sci.osaka-u.ac.jp, y\_nagoya@het.phys.sci.osaka-u.ac.jp, yamaguch@het.phys.sci.osaka-u.ac.jp}}\\
  \begin{abstract}
    We explore topological defects in the 4-dimensional pure $\mathbb{Z}_2$ lattice gauge theory.  
    This theory has 1-form $\mathbb{Z}_{2}$ center symmetry as well as the Kramers-Wannier-Wegner (KWW) duality.  
    We construct the KWW duality topological defects in the similar way to that constructed by Aasen, Mong, Fendley \cite{Aasen:2016dop} for the 2-dimensional Ising model.  
    These duality defects turn out to be non-invertible.  We also construct the 1-form $\mathbb{Z}_{2}$ symmetry defects as well as the junctions among KWW duality defects and 1-form $\mathbb{Z}_{2}$ center symmetry defects.  
    The crossing relations among these defects are derived.
    The expectation values of some configurations of these topological defects are calculated by using these crossing relations.
  \end{abstract}
\end{center}
\tableofcontents
\vspace{4ex}
\section{Introduction}
\label{sec:Intro}
Recently, there has been progress in extending the concept of symmetry, 
and its application to the analysis of non-perturbative dynamics of quantum field theories.
One of them is so-called ``non-invertible symmetry.'' In 2 dimensions, such non-invertible symmetries are described by ``fusion categories'' and investigated actively in many papers including \cite{Feiguin:2006ydp,Carqueville:2012dk,Brunner:2013xna,Bhardwaj:2017xup,Chang:2018iay,Freed:2018cec,Lin:2019hks,Thorngren:2019iar,Komargodski:2020mxz,Huang:2021ytb,Inamura:2021wuo,Thorngren:2021yso,Sharpe:2021srf}.
Non-invertible symmetries in higher dimensions are less understood than those in 2 dimensions.
A class of non-invertible symmetries in higher dimensions is described by ``fusion $n$-categories'' \cite{Douglas:2018,Johnson-Freyd:2020}.
There have been several studies of 4-dimensional non-invertible symmetries and their applications \cite{Ji:2019jhk,Kong:2020cie,Rudelius:2020orz,Johnson-Freyd:2020twl,Heidenreich:2021tna}.
There have also been other studies of non-invertible symmetries and their applications including \cite{Kapustin:2010if,Nguyen:2021yld,Nguyen:2021naa,Delmastro:2021otj,Kong:2021equ}.

These symmetries and generalized symmetries are described by topological defects \cite{Gaiotto:2014kfa}.
An interesting approach is to construct such topological defects in lattices, initiated by Aasen, Mong, and Fendley (AMF) \cite{Aasen:2016dop,Aasen:2020jwb}.
One of the advantages of the AMF approach is that this construction is perfectly explicit.
In this paper, we employ this AMF approach to construct topological defects in higher dimensions.

In this paper, we study the 4-dimensional pure $\Zb_2$ lattice gauge theory \cite{Wilson:1974sk}.  We prepare the 4-dimensional cubic lattice and assign a variable $U_m=\pm1$ to each link $m$. The partition function of this theory is given, with a positive constant parameter $K$, by
\begin{align}
	Z=\sum_{\{U\}}\exp\qty(K\sum_{i\in P}\prod_{m\in\Box_i}U_m).
  \label{partitionfunction0}
\end{align}
Here $P$ is the set of all plaquettes, and $\Box_i$ is the set of four links included in the plaquette $i$.

Wegner \cite{Wegner:1984qt} has discovered a duality in this 4-dimensional pure $\Zb_2$ lattice gauge theory that is similar to the Kramers-Wannier duality \cite{kramers1941statistics,kramers1941statistics2} in the 2-dimensional Ising model.
Let us call this duality ``the KWW duality.''
According to the KWW duality, the 4-dimensional pure $\Zb_2$ lattice gauge theories with the parameter $K$ and $\hat{K}$ are equivalent
\footnote{Precisely speaking, the theory with parameter $K$ is equivalent to the theory with parameter $\hat{K}$ in which the 1-form $\Zb_2$ center symmetry is topologically gauged.
Such gauging does not affect the local dynamics. For example, the critical point is exact if there is only one phase transition.}
when they satisfy the relation
\begin{equation}
  \sinh{2K}\sinh{2\hat{K}}=1.
\end{equation}
This theory is in the confinement phase when $K$ is small, and in the deconfinement phase when $K$ is large.  If there is only one phase transition, one can conclude that the phase transition occurs at the self-dual point $K=\hat{K}=:K_c$ where
\begin{align}
	K_c=-\frac{1}{2}\log(-1+\sqrt{2}).
\end{align}
It is believed that this is actually the case, and the phase transition is first-order \cite{Creutz:1979kf,Creutz:1979zg}.

Another important symmetry in this theory is the 1-form $\Zb_2$ center symmetry \cite{Gaiotto:2014kfa}.  Wilson loops are charged under this center symmetry.

In this paper, we investigate the KWW duality and the 1-form $\Zb_2$ center symmetry in the pure $\Zb_2$ lattice gauge theory in 4 dimensions, by the AMF approach.  We explicitly construct duality defects and show that they are non-invertible. In particular, we find that the expectation value of a duality defect with a topology of $S^3$ is $\frac{1}{\sqrt{2}}$, not $1$, which implies the duality defect is non-invertible.  We also construct $\Zb_2$ symmetry defects by the AMF approach.

Furthermore, we investigate the crossing relations among duality defects and $\Zb_2$ symmetry defects.
The crossing relations are not closed within duality defects.  It is closed only when $\Zb_2$ symmetry defects are included.
To consider these relations, we introduce defect junctions. 
In particular, we derive the crossing relations between two configurations of duality defects which are related by cutting and gluing solid tori.
By using junctions and these relations, we can calculate the expectation values of the duality defects with various topologies.
These values are invariants of embeddings of 3-dimensional manifolds into 4 dimensions because duality defects are topological defects.
We compute expectation values of $S^3$ and $S^1\times S^2$.

The construction of this paper is as follows.
In Sec.~\ref{sec:4dZ2}, we explain the setup of the pure $\Zb_2$ lattice gauge theory in 4 dimensions. In Sec.~\ref{sec:topologicaldefects}, we construct duality defects and $\Zb_2$ symmetry defects by AMF approach.  
We solve the commutation relations that are satisfied by the topological defects, and find the solution.  
We show that duality defects are non-invertible by calculating their expectation values.  
We also construct $\Zb_2$ symmetry defects and their junctions with duality defects.  
Then we find crossing relations among duality defects and 
$\Zb_2$ symmetry defects.  Sec.~\ref{sec:conclusion} is devoted to the conclusions and discussions.

\section{4-dimensional \texorpdfstring{$\Zb_2$}{Z2} lattice gauge theory}
\label{sec:4dZ2}
In this section, we explain our formulation of the 4-dimensional $\Zb_2$ lattice gauge theory.

\begin{figure}[H]
  \centering
  \includegraphics[width=5cm]{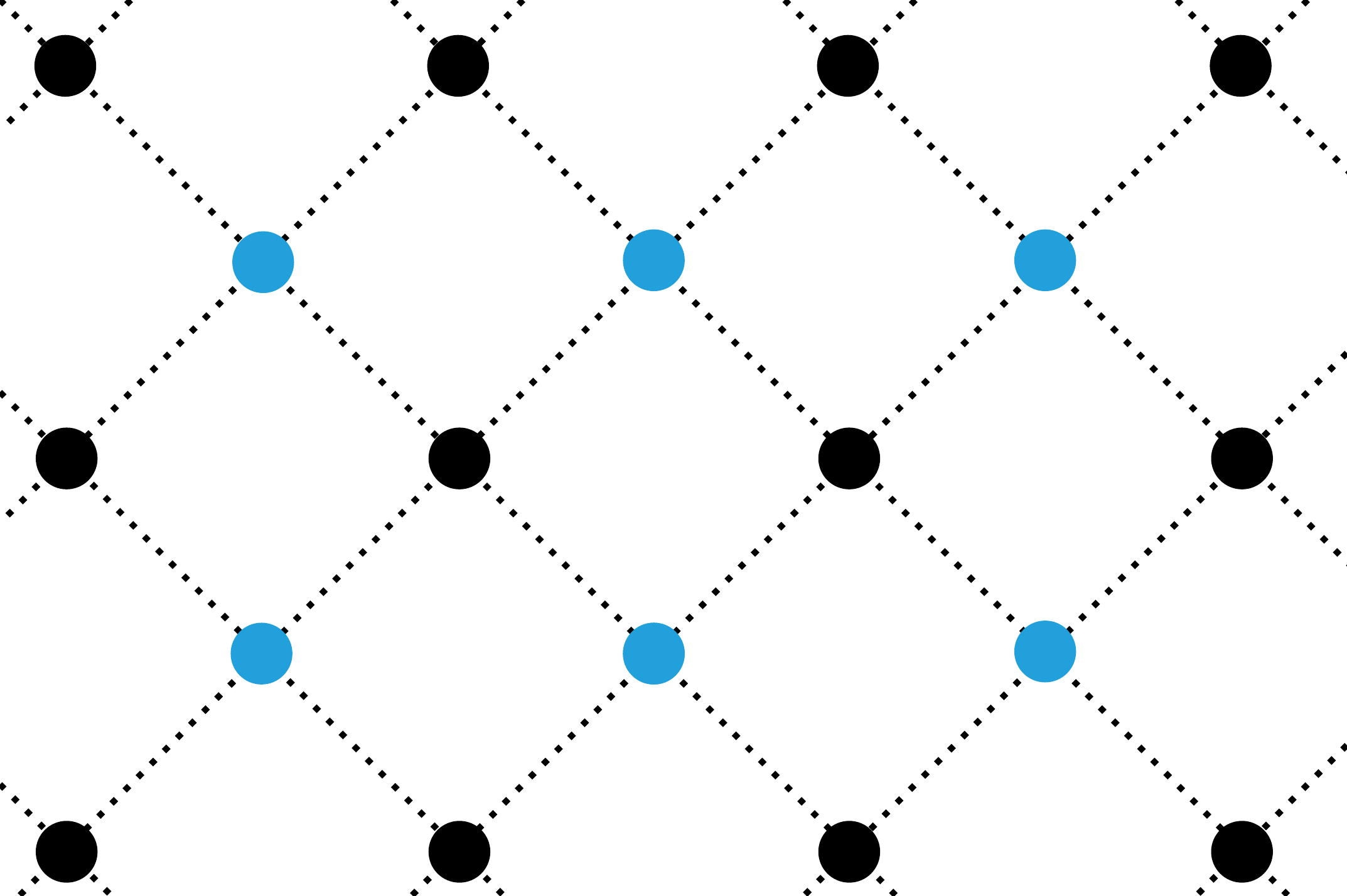}
  \caption{A schematic illustration of lattices. Although it is depicted as 2-dimensional in this figure, the actual lattices treated in this paper are 4-dimensional ones. The black lattice represents the lattice $\Lambda$, and the blue lattice represents the lattice $\hat{\Lambda}$. They are dual to each other.}
  \label{fig:lattice}
\end{figure}

We introduce two kinds of lattices in order to consider duality defects later as shown in Figure \ref{fig:lattice}. Our formulation in 4 dimensions is almost parallel to \cite{Aasen:2016dop} in 2 dimensions.

To be more specific, we introduce coordinates $(x_1,x_2,x_3,x_4)$ of $\Rb^{4}$. 
The coordinates of points in $\Lambda$ are even integers, while those of $\hat{\Lambda}$ are odd integers.
In other words, we define $\Lambda:=\{(x_1,x_2, x_3,x_4)|x_1,x_2,x_3,x_4\in 2\Zb\}$, and $\hat{\Lambda}:=\{(x_1,x_2,x_3,x_4)|x_1,x_2,x_3,x_4\in 2\Zb+1\}$.  
We call the line segment connecting the nearest two points in each lattice a link.
We also call the smallest square consisting of such links a plaquette.
The line segment connecting a point on $\Lambda$ and a point on $\hat{\Lambda}$ is not called a link.

We assign the link variables $U_m$ as explained in Sec.~\ref{sec:Intro} to the links in $\Lambda$, while we do not assign link variables to the links in $\hat{\Lambda}$. 
For this reason, we refer to the lattice $\Lambda$ to which the link variable is assigned as the active lattice, and sites, links, and plaquettes on it as active sites, active links, and active plaquettes, respectively. 
On the other hand, the lattice $\hat{\Lambda}$, to which no link variables are assigned, is called the inactive lattice, and links, sites, and plaquettes on it are called inactive sites, inactive links, and inactive plaquettes, respectively.

\begin{figure}[H]
  \centering
  \includegraphics[width=7cm]{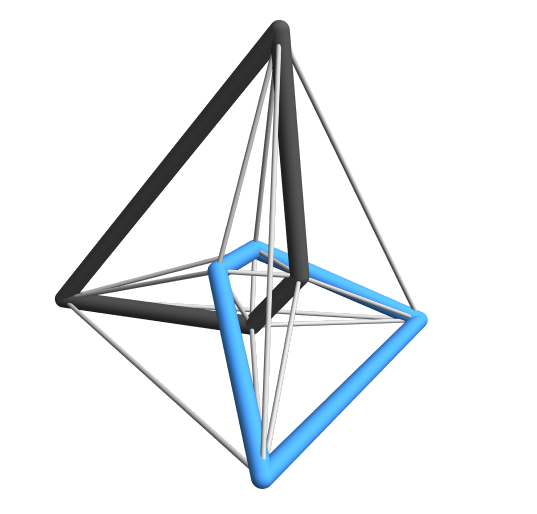}
  \caption{Stereographic projection of a 16-cell into 3 dimensions.  The black plaquette represents an active plaquette, and the blue plaquette represents an inactive plaquette. The 16-cell consists of 16 tetrahedrons, each of which contains one active link and one inactive link.}
  \label{fig:16}
\end{figure}

We can regard the basic unit of such a lattice system as a 16-cell (see for example \cite{coxeter2012regular,wikipedia16-cell}). 
As shown in Figure \ref{fig:16}, the 16-cell consists of an active plaquette and an inactive plaquette that share a center. 
The 16-cell consists of 16 tetrahedrons each of which contains an active link and an inactive link. The surface of the 16-cell is homeomorphic to $S^3$.

Notice that there is a one-to-one correspondence between active plaquettes and 16-cells. 
For example, let us choose an active plaquette $p$ composed of four points $(0,0,0,0),(0,0,0,2),(0,0,2,0),(0,0,2,2)$ in $\Lambda$. 
The center of this plaquette is $(0,0,1,1)$. 
Then the inactive plaquette $\tilde{p}$ composed of four points $(\pm1,\pm1,1,1)$ in $\hat{\Lambda}$ shares the same center as $p$. 
There is a 16-cell containing $p$, $\tilde{p}$ and these 8 points.

Since there is a one-to-one correspondence between 16-cells and active plaquettes, assigning the Boltzmann weights to the active plaquettes is equivalent to assigning the Boltzmann weights to the 16-cells. Let $a_i=0,1$, $(i=1,2,3,4)$ be the link variables assigned to the four active links in a 16-cell. We define the Boltzmann weight to this 16-cell as
\begin{equation}
    W(a_1,a_2,a_3,a_4)=\exp(K(-1)^{(a_1+a_2+a_3+a_4)}).
    \label{boltzmannweight16cell}
\end{equation}
This definition is equivalent to the Boltzmann weight in the $\Zb_2$ gauge theory in Eq.~\eqref{partitionfunction0} if the link variables for the link $m$ are identified as $U_{m}=(-1)^{a_m}$.

It is convenient to introduce constant weights for links and sites. 
We define the weights for active links, active sites, inactive links, and inactive sites as $s,\ l,\ \Bar{s}$, and $\Bar{l}$, respectively.
These weights are required to define the duality defect later. They are determined in Sec.~\ref{sec:KWdefect}.

We define the partition function of this theory as
\begin{align}
    Z=\sum_{\{a\}}\qty(\prod_{\substack{ \text{active}\\ \text{sites}}}s)\qty(\prod_{\substack{ \text{active}\\ \text{links}}}l)\qty(\prod_{\substack{ \text{inactive}\\ \text{sites}}}\Bar{s})\qty(\prod_{\substack{ \text{inactive}\\ \text{links}}}\Bar{l})\prod_{i\in C}W(a_{j_1(i)},a_{j_2(i)},a_{j_3(i)},a_{j_4(i)}),
\end{align}
where $C$ is the set of all 16-cells, and $j_1(i),\ j_2(i),\ j_3(i),$ and $j_4(i)$ are the four active links in the 16-cell $i$. $a_j$ is the link variable assigned to the active link $j$.
Besides the constant normalization, it is identical to the partition function \eqref{partitionfunction0}.

\section{Topological defects}
\label{sec:topologicaldefects}
In this section, we discuss topological defects and junctions among them in the $\Zb_2$ lattice gauge theory following AMF approach \cite{Aasen:2016dop,Aasen:2020jwb}. 
We require the defect commutation relations of topological defects and their junctions, and determine their weights.
We derive crossing relations among them.
We calculate the expectation values of some configurations of defects by using these crossing relations.
In particular, we find that the defect associated to the KWW duality is non-invertible.

\subsection{Duality defect}
\label{sec:KWdefect}
In this subsection, we discuss duality defects in the 4-dimensional $\Zb_2$ pure gauge theory. Duality defects are 3-dimensional operators supported on the boundary between two regions associated by the KWW duality. The active lattice and the inactive lattice are swapped across a duality defect, as a KW defect in the 2-dimensional Ising model \cite{Aasen:2016dop}.

\begin{figure}[htbp]
  \centering
  \includegraphics[width=5cm]{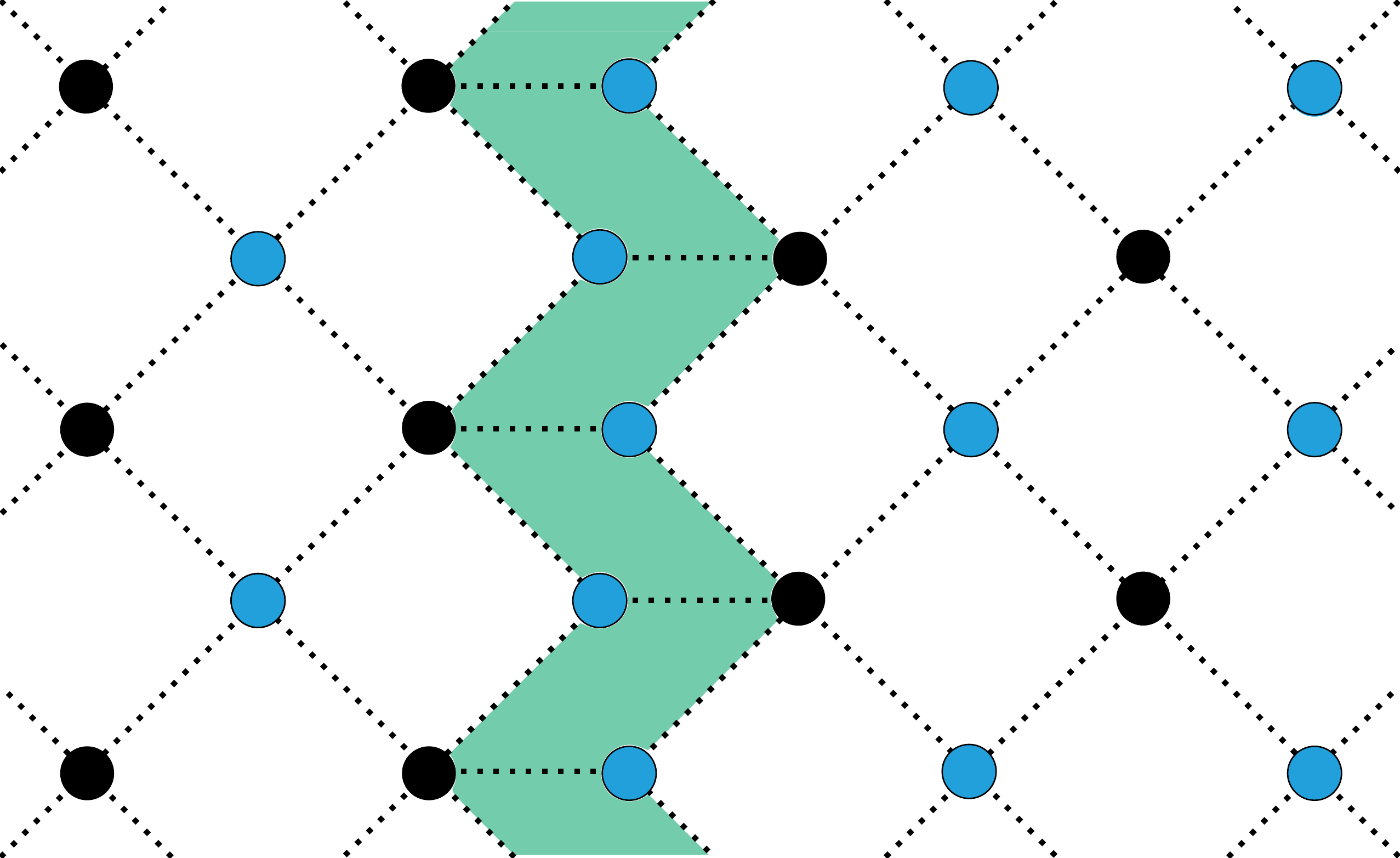}
  \caption{A schematic illustration of the duality defect. The black dots represent the active lattice and the blue dots represent the inactive lattice. A duality defect is located at the boundary between the two regions. The active lattice (black dots) and the inactive lattice (blue dots) are swapped across the duality defect. A unit cell of the duality defect is a tetrahedral prism depicted as a green parallelogram in this figure.}
  \label{fig:duality_defect1}
\end{figure}

\begin{figure}[htbp]
  \centering
  \includegraphics[width=5cm]{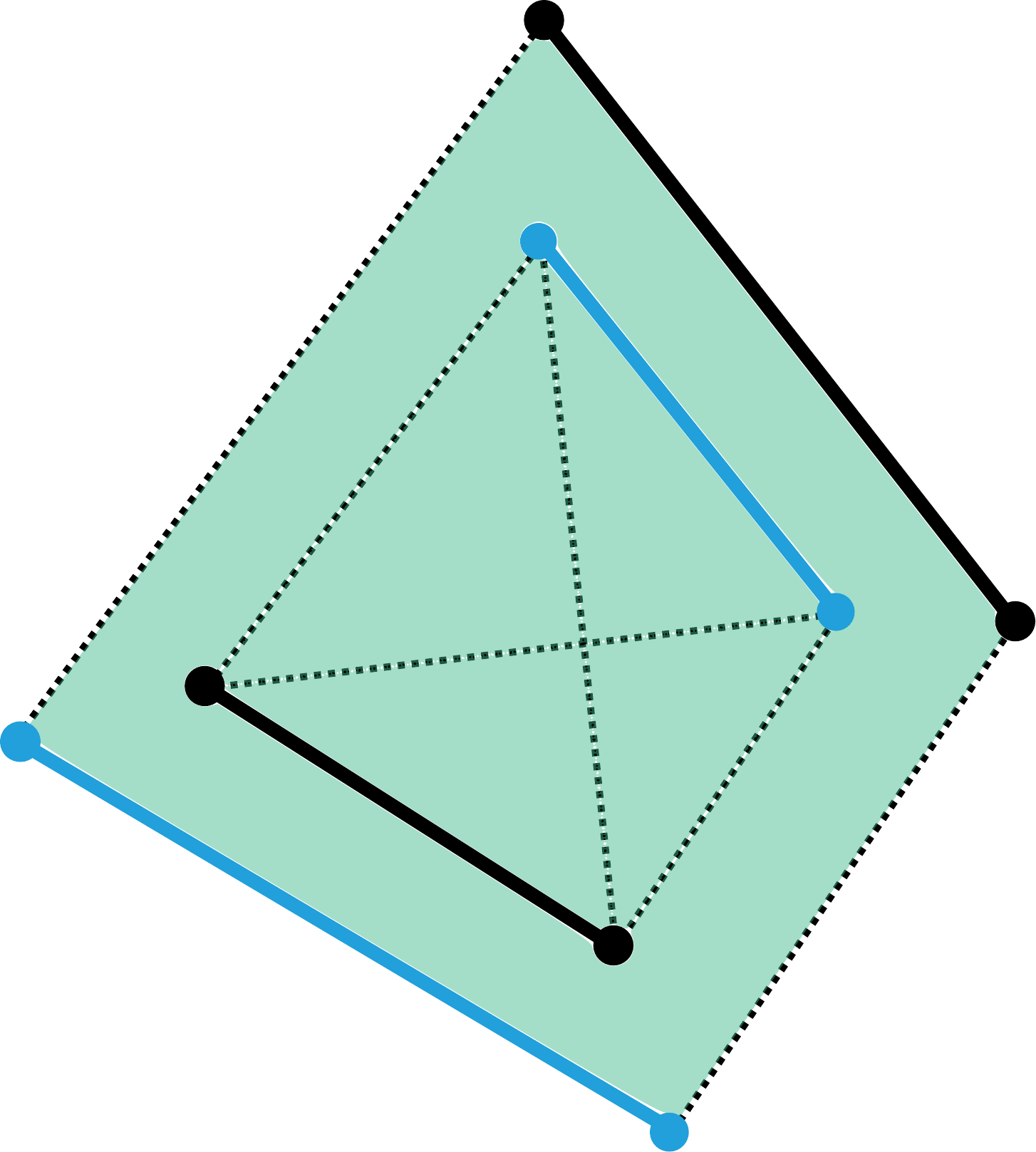}
  \caption{The building block of duality defects in the 4-dimensional $\Zb_2$ lattice gauge theory.  The 3-dimensional surface is composed of tetrahedrons each of which includes an active link and an inactive link.  A tetrahedron on which a duality defect is supported is doubled and becomes a tetrahedral prism. An inactive link is put on the edge of a tetrahedron in this tetrahedral prism associated to the active link of the other tetrahedron and vice versa. }
  \label{fig:duality_defect2}
\end{figure}

The 3-dimensional unit cell on the tessellation by regular 16-cells is a regular tetrahedron. 
Therefore, we employ these tetrahedrons as the building blocks of the 3-dimensional surface on which a duality defect is supported. 
In our model, each such tetrahedron consists of an active link and an inactive link. 
The active lattice and the inactive lattice are swapped across the duality defect. 
In order to implement this property, it is convenient to double each tetrahedron on which the duality defect is supported, and swap the active link and the inactive link (see Figure \ref{fig:duality_defect1}).  
As a result, the building block of duality defects is a tetrahedral prism (see Figure \ref{fig:duality_defect2}).

We want to construct topological duality defects.
We expect that such topological duality defects exist because the duality transformation just changes the description of the theory, but it does not change the observables.
In the following, we impose the defect commutation relations so that the duality defect is topological, and find a solution.

\begin{figure}[htbp]
  \centering
  \includegraphics[scale=0.3]{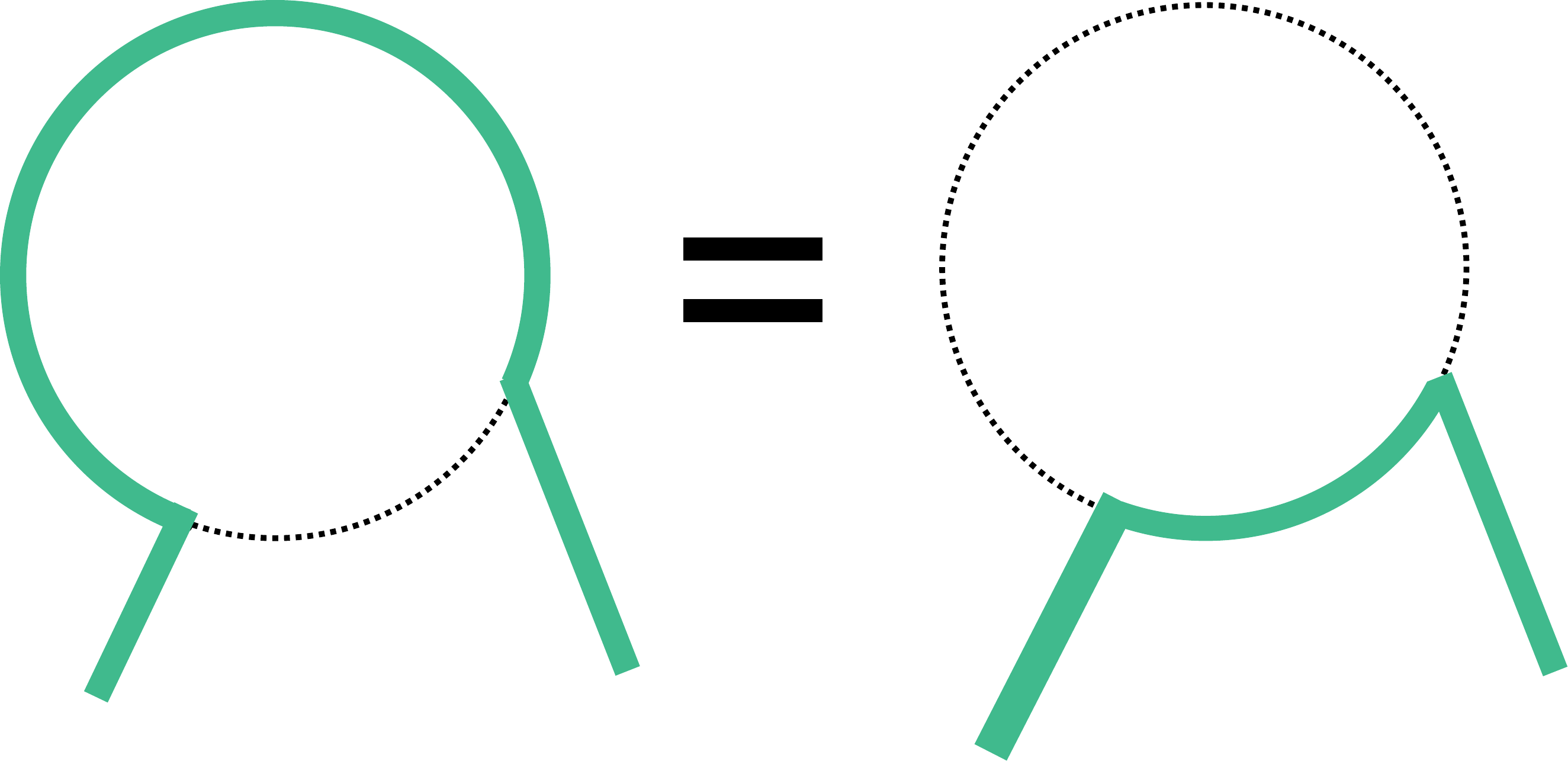}
  \caption{A schematic illustration of a defect commutation relation. 
  The circle represents a 16-cell and a green line represents a duality defect. 
  The defect commutation relation implies that the value of the duality defect remains the same even if it is deformed without changing the topology.}
  \label{fig:commu}
\end{figure}

Now we focus on a single 16-cell in our setup and consider defect commutation relations (see Figure \ref{fig:commu}).
There are sixteen tetrahedrons on the surface of the 16-cell.
We consider a configuration of a duality defect \confA.
Some tetrahedrons out of 16 are filled by this defect \confA and the others are not.
We also consider a deformed configuration \confB in which the tetrahedrons which are not filled by \confA is filled and vice versa.
\confA and \confB are the same away from the focused 16-cell.
We require \confA and \confB have the same weight if \confA and \confB have the same topology.
Such an equality between two different duality defect configurations is called a defect commutation relation.

The topology of \confA and \confB are the same without any ambiguity if and only if the filled tetrahedrons on the focused 16-cell satisfy the following conditions.
\begin{itemize}
\item The filled tetrahedrons on the 16-cell of \confA and \confB are both non-empty sets.
\item The configurations \confA and \confB restricted on the surface of the focused 16-cell are both simply-connected.
\item There is no ambiguity in the above condition. In other words, there is no connection such that the duality defects are connected only by sites or links. 
\end{itemize}

We explain the defect commutation relations in more detail. 
Each building block of a duality defect contains two active links. 
Let us take a building block and let the link variables assigned to these two active links $a,\tilde a=0,1$. 
Then we assign a weight $D(a,\tilde a)$ to this building block. 
We also define the weights of each component of the active lattice and the inactive lattice as in Sec.~\ref{sec:4dZ2}.  
We denote the weight of an active link, the weight of an active site, the weight of an inactive link, and the weight of an inactive site as $l,s,\bar{l} ,\bar{s}$, respectively.

Let us choose a 16-cell in our system away from defects and explain a bit more on the notations.
Let the labels of the active links in this 16-cell be $m=1,2,3,4$ and the labels of the inactive links be $\tilde n=\tilde1,\tilde2,\tilde3,\tilde4$.
We also denote the set of four active links in the 16-cell as $M=\{1,2,3,4\}$ and the set of four inactive links as $\tilde N=\{\tilde1,\tilde2,\tilde3,\tilde4\}$.
A pair of an active link and an inactive link $(m,\tilde n)$ can be used to specify a tetrahedron in the 16-cell.
Let $U$ be the set of all tetrahedrons
\begin{align}
U=\{(m,\tilde n)|m=1,2,3,4,\ \tilde n=\tilde1,\tilde2,\tilde3,\tilde4\}.\label{eq:tetrahedrons}
\end{align}
The weight of this 16-cell is the Boltzmann weight $W(a_1,a_2,a_3,a_4)$ in Eq.~\eqref{boltzmannweight16cell}. Here, $a_m$ is the link variable assigned to the active link $m$.

Next, we consider a 16-cell with duality defects on its surface in the configuration \confA. Let $I\subset U$ be the set of tetrahedrons filled by the duality defects in the configuration \confA.
Since the duality defect is defined by doubling the tetrahedron, we have to take active links coming from this doubling into account, in addition to the set of active links $M$ contained within this 16-cell. 
These additional active links are the counterparts of the inactive links included in tetrahedrons contained in the set $I$.   
We can use $\tilde n$, which is originally the label of an inactive link in the 16-cell, as the label of such an additional active link.
Thus, the set of all such additional active links is $\tilde E=\{\tilde n|(m,\tilde n)\in I\}$. Also, let $\tilde a_{\tilde n}$  the link variable of the additional active link $\tilde n$.  By using these notations, we can write down the weight of each building block of this duality defect located on the surface of the 16-cell as 
\begin{align}
D(a_m,\tilde a_{\tilde n}),\quad (m,\tilde n)\in I.
\end{align}
Then the total weight of the 16-cell and the duality defect located on the 16-cell in the configuration \confA is written as
\begin{align}
W(a_1,a_2,a_3,a_4)\prod_{(m,\tilde n)\in I} D(a_m,\tilde a_{\tilde n}).
\end{align}

The defect commutation relations are conditions that the weight of the configuration \confA before the deformation of Figure \ref{fig:commu} with the defect placed on $I$ are equal to the weight of the configuration \confB  after the deformation with the defect placed on its complement $\bar I =U \setminus I$.
Therefore, we should also consider the configuration \confB.
One should notice that the 16-cell and its active links in the configuration \confB are included in the region of the other side of the defect where the active lattice and the inactive lattice are swapped.
In order to compare it to the weights of the configuration \confA, it is more convenient to denote the labels of the active links within the 16-cell after the deformation by $\tilde n=\tilde1,\tilde2,\tilde3,\tilde4$, and the labels of the inactive links within the 16-cell by $m=1,2,3,4$.  The link variable for $\tilde n$ is denoted by $\tilde a_{\tilde n}$ since they couple to the other part of the system just the same way as $\tilde a_{\tilde n}$ in the configuration \confA if exist.  In this notation, the Boltzmann weight for the 16-cell is $W(\tilde a_{\tilde 1},\tilde a_{\tilde 2},\tilde a_{\tilde 3},\tilde a_{\tilde 4})$.  The total weight of the 16-cell and the duality defects located on the 16-cell in the configuration \confB is written as 
\begin{align}
W(\tilde a_{\tilde 1},\tilde a_{\tilde 2},\tilde a_{\tilde 3},\tilde a_{\tilde 4})\prod_{(m,\tilde n)\in \bar I} D(a_m,\tilde a_{\tilde n}).
\end{align}
We also consider $E=\{m|(m,\tilde n)\in \bar I\}$. This is the set of inactive links $m$ touching the duality defect located on the tetrahedrons in $\bar I$. Each of these inactive links is doubled by the duality defect and corresponds to an additional active link.  
The link variable $a_m$ for the active link $m$ couples to the other part of the system just the same way as that of the link variable $a_m$ in the configuration \confA.
Therefore, they are identified with each other.

 We impose the following defect commutation relations as conditions for $D(a,\tilde a),l,s,{\bar l},{\bar s},\ K$ .
\begin{align}
\label{eq;com relation}
\sum_{M\setminus E} W(a_1,a_2,a_3,a_4)s^{\alpha_1}l^{\beta_1}{\bar s}^{\tilde {\alpha_1}}{\bar l}^{\tilde {\beta_1}}\prod_{(m,\tilde n)\in I}D(a_m,\tilde a_{\tilde n}) = \sum_{\tilde N\setminus \tilde E} W(\tilde a_{\tilde 1},\tilde a_{\tilde 2},\tilde a_{\tilde 3},\tilde a_{\tilde 4})s^{\alpha_2}l^{\beta_2}{\bar s}^{\tilde {\alpha_2}}{\bar l}^{\tilde {\beta_2}}\prod_{(m,\tilde n)\in \tilde I}D(a_m,\tilde a_{\tilde n}).
\end{align}
Here the number of active sites, active links, inactive sites, and inactive links in the configuration \confA are denoted by $\alpha_1,\beta_1,\tilde {\alpha_1}$, and $\tilde {\beta_1}$, respectively.
The number of active sites, active links, inactive sites, and inactive links on the configuration \confB are denoted by $\alpha_2,\beta_2,\tilde {\alpha_2},$ and $\tilde {\beta_2}$, respectively.  
Also, the sums $\sum_{M\setminus E}, \sum_{\tilde N\setminus \tilde E} $ are defined by 
\begin{align}
\sum_{M\setminus E}:=\prod_{m\in M\setminus E}\sum_{a_m=0,1},\qquad
\sum_{\tilde N \setminus \tilde E}:=\prod_{\tilde n\in \tilde N\setminus \tilde E}\sum_{\tilde a_{\tilde n}=0,1}.
\end{align}

In Eq.~\eqref{eq;com relation}, on the left-hand side we sum over the link variables for the links in $M\setminus E$, and on the right-hand side we sum over the link variables for the links in $\tilde N\setminus \tilde E$.
Let us explain a little bit more on this summation.
To begin with, both sides of the defect commutation relations are the contributions to the partition function or correlation functions with the defects from the 16-cell and the building blocks of the defect on the surface of it.
The defect commutation relations are required to be satisfied for arbitrary link variables for the active links that are included in other 16-cells where some other operator may be inserted.
On the other hand, the active links contained in $M\setminus E,\ \tilde N\setminus \tilde E$ are not contained in other 16-cells.
These are the degrees of freedom that arise and disappear before and after the deformation.  
Therefore, the sums inherited from the partition function are taken for the link variables for such links.

We determine the values $D(a,\tilde a),l,s,{\bar l},{\bar s},K$ by solving these defect commutation relations.
In a physically sensible solution, these values satisfy
\begin{align}
  D(a,\tilde a)\ne 0,\quad l,s,{\bar l},{\bar s} >0,\quad K\in \Rb,\quad K\ne 0.
\end{align}
There is a unique physically sensible solution up to the sign of $D(a,\tilde a)$.  
The solution is
\footnote{We have used Mathematica to find the solution.  We have learned this method from Kantaro Ohmori's lecture ``Categorical symmetry in $1+1$ dimensions'' (in Japanese) in CREST online workshop ``Theoretical studies of topological phases of matter.'' We would like to thank him for this excellent lecture.}
{\allowdisplaybreaks[1]
\begin{align}
D(a,\tilde a)=\ \ &\includegraphics[width=2cm,height=20mm,align=c]{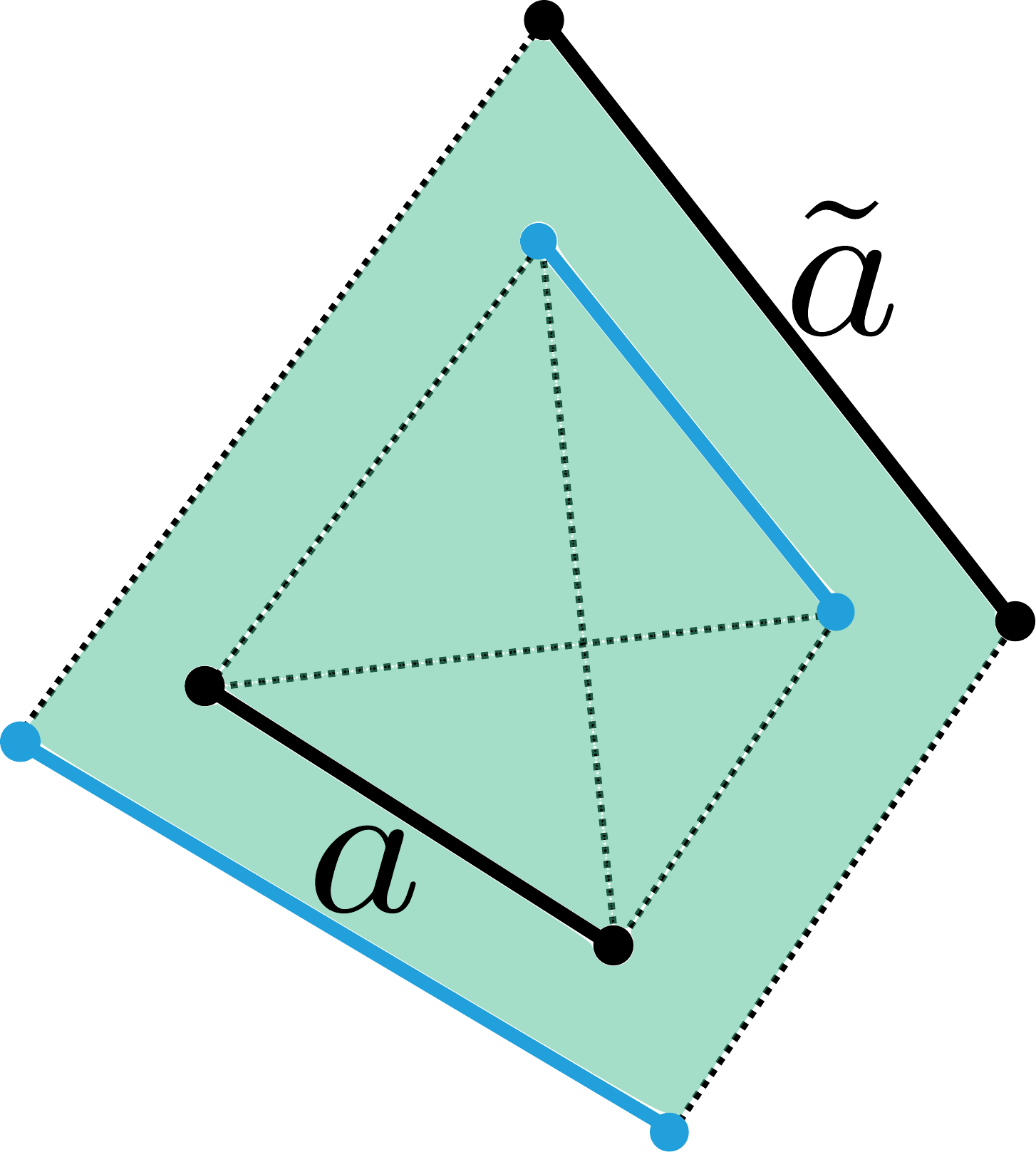}\ \ =(-1)^{a{\tilde a}},&&\label{solution1-1}\\
l=\ \ &\includegraphics[width=3cm,height=3mm,align=c]{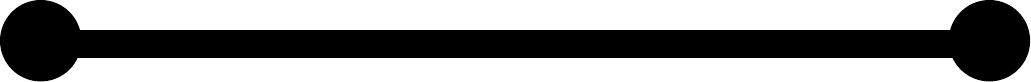}\ =\frac{1}{\sqrt{2}},\qquad &&s=\ \includegraphics[width=3mm,height=3mm,align=c]{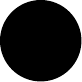}\ =\frac{1}{\sqrt{2}},\label{weightactivelink}\\
{\bar l}=\ \ &\includegraphics[width=3cm,height=3mm,align=c]{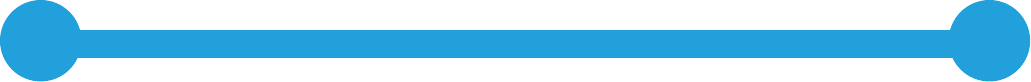}\ =1,\qquad
&&{\bar s}=\ \includegraphics[width=3mm,height=3mm,align=c]{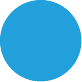}\ =1,\\
K=&K_c=-\frac{1}{2}\log(-1+\sqrt{2}),\qquad &&W(a_1,a_2,a_3,a_4)=\exp(K(-1)^{(a_1+a_2+a_3+a_4)}). \label{solution1-2}
\end{align}
}
Notice that the value of $K$ is determined to be the critical value $K_c$. The sign ambiguity for $D(a,\tilde a)$ does not affect the observables at least locally.  We discuss a bit more on this sign ambiguity in Sec.~\ref{sec:conclusion}.

Let us show the gauge invarinance of our duality defects.  The building block by itself is not gauge invariant.
To see this, we consider the gauge transformation at a site $S$ in a building block $B$.
Let $L$ the active link including $S$ contained in $B$.
Also let $L'$ the other active link in $B$. 
$a$ and $\tilde{a}$ denote the link variables of $L$ and $L'$, respectively. 
Then the weight of this building block $B$ is transformed by this gauge transformation as 
\begin{align}
D(a,\tilde a)\ \ \rightarrow \ \ D(1-a,\tilde a)=(-1)^{(1-a){\tilde a}}.
\end{align}
So, it is not gauge invariant.
However, the whole of the duality defect is gauge invariant as shown below.
Since the whole of the duality defect does not have a boundary, there is a unique building block which includes $S$ and $L'$ and does not include $L$;
let $B'$ denote this building block.
Also, let $b$ the link variable of the active link including $S$ contained in $B'$.
Then the weights of these building blocks $B,B'$ are transformed by the gauge transformation at the active site $S$ as
\begin{equation}
D(a,\tilde a)D(b,\tilde a)
\ \ \rightarrow \ \ D(1-a,\tilde a)D(1-b,\tilde a)
=(-1)^{(1-a){\tilde a}}(-1)^{(1-b){\tilde a}}
=(-1)^{a{\tilde a}}(-1)^{b{\tilde a}}
=D(a,\tilde a)D(b,\tilde a).
\end{equation}
Therefore, this pair of building blocks $B,B'$ is invariant by this gauge transformation.
The building blocks including the site $S$ are divided into such pairs, so they are invariant.
The building blocks which do not include $S$ are trivially invariant.
Thus we can conclude that the whole of the duality defect is gauge invariant.

\begin{figure}[htbp]
  \centering
  \includegraphics[scale=0.3]{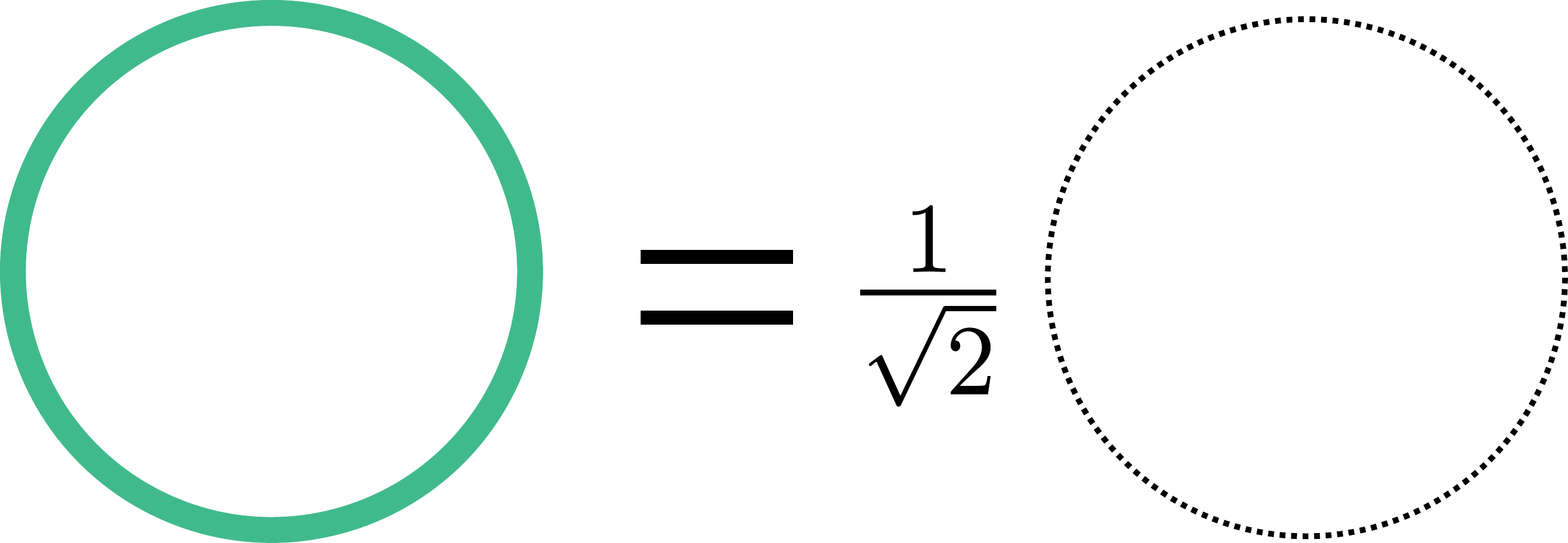}
  \caption{A schematic illustration of Eq.\eqref{eq;qdim}. The circles represent the 16-cell and the green line represents the duality defect. }
  \label{fig:qdim}
\end{figure}

Now we show that the duality defects constructed here are non-invertible.
The easiest way to do this is to consider the case where the set of filled tetrahedrons on the 16-cell in the configuration \confB is empty as shown in Figure \ref{fig:qdim}.
In the configuration \confA, the duality defect is located on all tetrahedrons of the 16-cell and forms a closed manifold $S^3$.
In this case, the deformation does change the topology of the defect from $S^3$ to empty, and therefore, the Eq.~\eqref{eq;com relation} is not required.
Instead, one finds that the following equation is satisfied by substituting our solution \eqref{solution1-1} -- \eqref{solution1-2}.
\begin{align}
\label{eq;qdim}
\sum_{a_1,a_2,a_3,a_4=0,1}W(a_1,a_2,a_3,a_4)s^{8}l^{8}{\bar s}^{8}{\bar l}^{8}\prod_{(m,\tilde n)\in U}D(a_m,\tilde a_{\tilde n}) = \frac{1}{\sqrt{2}}W(\tilde a_{\tilde 1},\tilde a_{\tilde 2},\tilde a_{\tilde 3},\tilde a_{\tilde 4})s^{4}l^{4}{\bar s}^{4}{\bar l}^{4}.
\end{align}
For a symmetry defect placed on a closed manifold without any operator insertion inside, the weight is identical with the empty configuration \cite{Gaiotto:2014kfa}.  However, for our duality defect they are not identical, but their ratio is $1/\sqrt{2}$.  As a result, we can conclude that our duality defects are non-invertible.

\subsection{\texorpdfstring{$\Zb_2$}{Z2} 1-form symmetry defects}
\label{sec:Z2defect}
In this subsection, we explain topological defects associated to the 1-form $\Zb_2$ center symmetry \cite{Gaiotto:2014kfa}.
The 4-dimensional $\Zb_2$ pure gauge theory has 1-form $\Zb_2$ center symmetry. 
By the general symmetry argument \cite{Gaiotto:2014kfa}, the symmetry defects associated to this symmetry are supported on 2-dimensional manifolds.
The charged objects are Wilson loops.
A $\Zb_2$ symmetry defect acts on the Wilson loop by linking it.
The sign of the expectation value including the Wilson loop is flipped by the action of the $\Zb_2$ symmetry defect.

\begin{figure}[H]
  \centering
  \includegraphics[width=6cm]{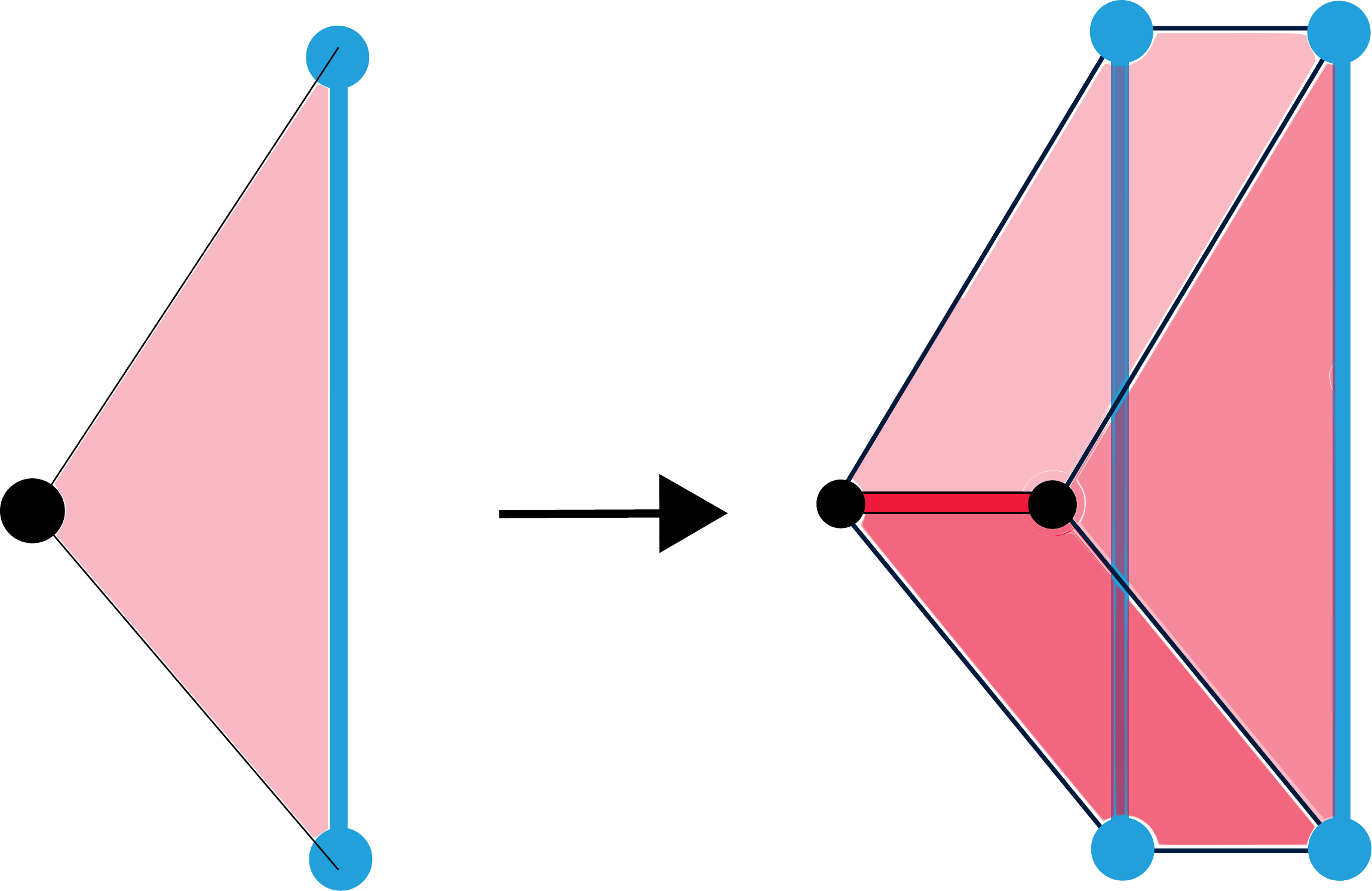}
  \caption{The building block of the 1-form $\Zb_2$ center symmetry defect. 
  A black dot represents the midpoint of an active link.  A blue dot represents an inactive site and a blue line represents an inactive link. 
  A building block of the surface on which a 1-form $\Zb_2$ symmetry defect is supported is a triangle formed by an inactive link and the midpoint of an adjacent active link. 
  A triangle on which a $\Zb_2$ symmetry defect is supported is doubled and becomes a triangular prism. }
  \label{fig:Z2}
\end{figure}

\begin{figure}[H]
  \centering
  \includegraphics[width=10cm]{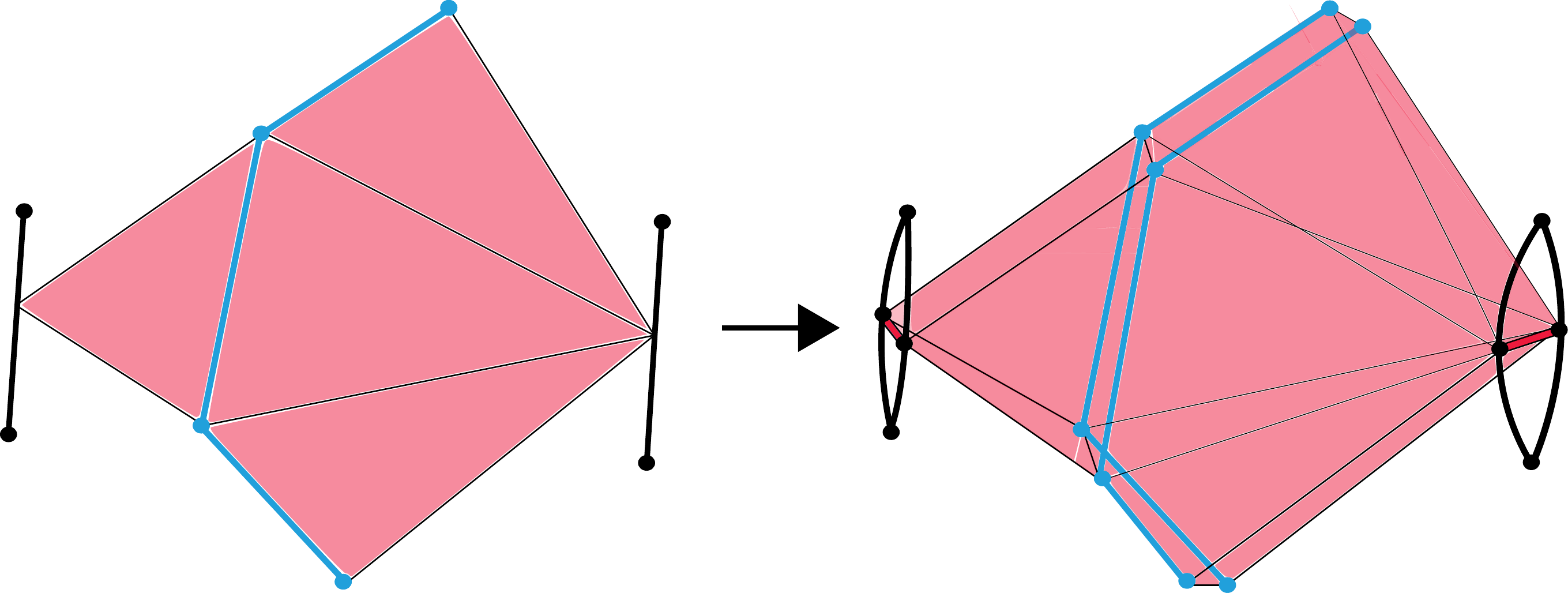}
  \caption{The surface is made by connecting several building blocks of the 1-form $\Zb_2$ symmetry defect. A black line represents an active link. A blue dot represents an inactive site and a blue line represents an inactive link. }
   \label{fig:Z2net}
\end{figure}

Since $\Zb_2$ symmetry defects are not codimension 1, the construction of the symmetry defects is slightly different from that of duality defects explained in Sec.~\ref{sec:KWdefect} and the defects constructed in \cite{Aasen:2016dop} and \cite{Aasen:2020jwb}.
We consider a closed 2-dimensional surface constructed by the triangles each of which is formed by an inactive link and the midpoint of an adjacent active link. 
We consider a $\Zb_2$ symmetry defect supported on it. 
We deform the lattice to describe these $\Zb_2$ symmetry defects as follows.
An inactive link and the inactive sites contained in the $\Zb_2$ symmetry defects are doubled.
On the other hand, if the midpoint of an active link is contained in the $\Zb_2$ symmetry defect, the active link is doubled while the two active sites at the endpoints of the active link are not doubled. 
Each triangle on this surface is doubled and becomes a triangular prism by this deformation (see Figures \ref{fig:Z2}, \ref{fig:Z2net}). 
We employ this triangular prism as a building block of $\Zb_2$ 1-form symmetry defects.

We assign weights to elements of $\Zb_2$ symmetry defects as follows.  A building block contains the midpoints of two active links.  The link variables of these two active links are denoted by $b,c=0,1$.
We want to define the 1-form $\Zb_2$ symmetry defect in such a way that their action flips the sign of a Wilson loop.
Therefore, we assign the weight $Z_2(b,c)=\sigma^x_{b,c}=(1-\delta_{b,c})$ to each building block.
We also assign a weight $z=\sqrt{2}$ to each pair of doubled active links so that the extra weight \eqref{weightactivelink} for the doubled active link is canceled.  The weights for these elements are summarized as follows.

\begin{align}
Z_2(b,c)=\ \ \includegraphics[width=1.5cm,height=20mm,align=c]{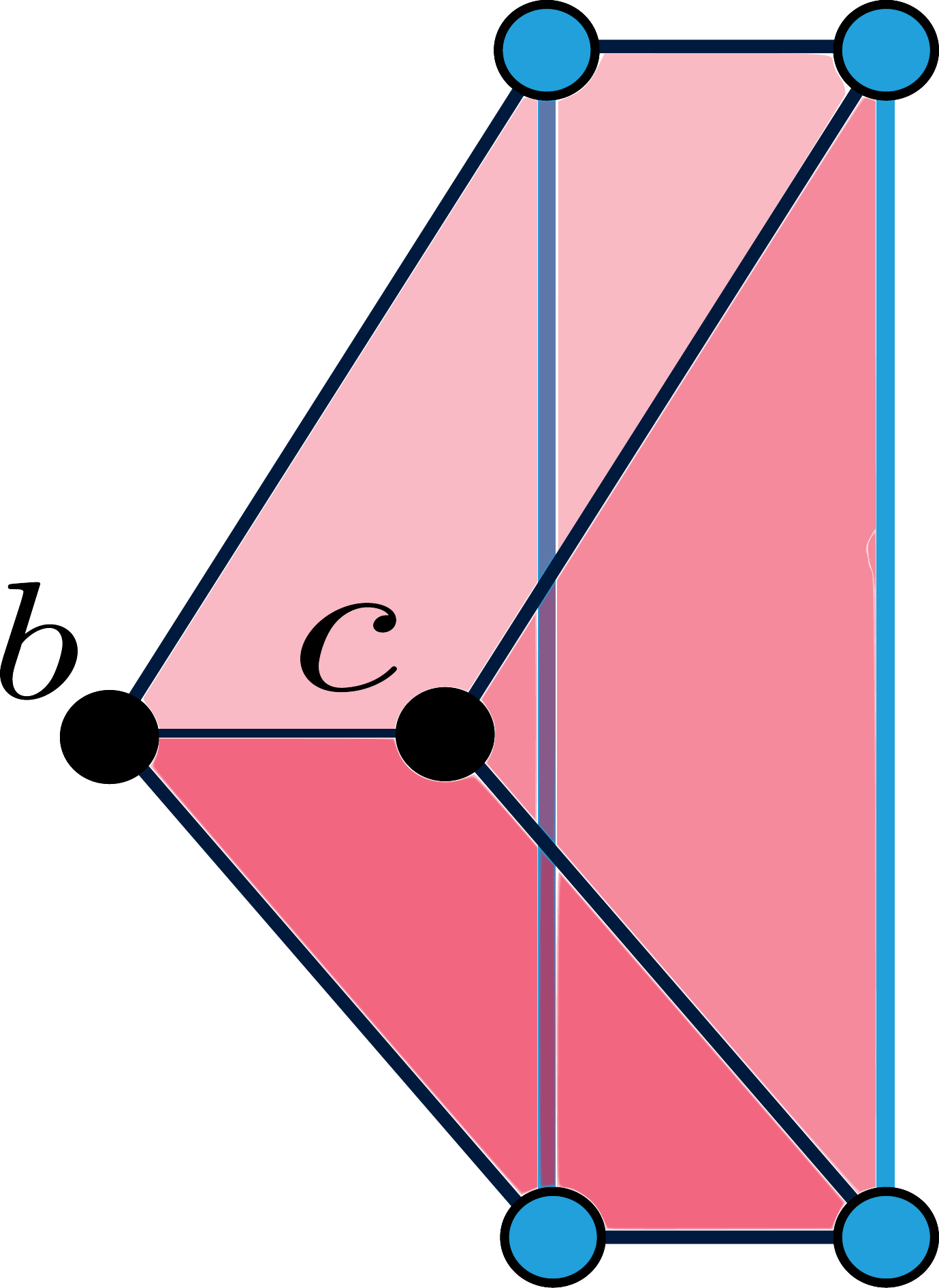}\ \ =\sigma^x_{b,c},\qquad z=\includegraphics[width=3cm,%
align=c
]{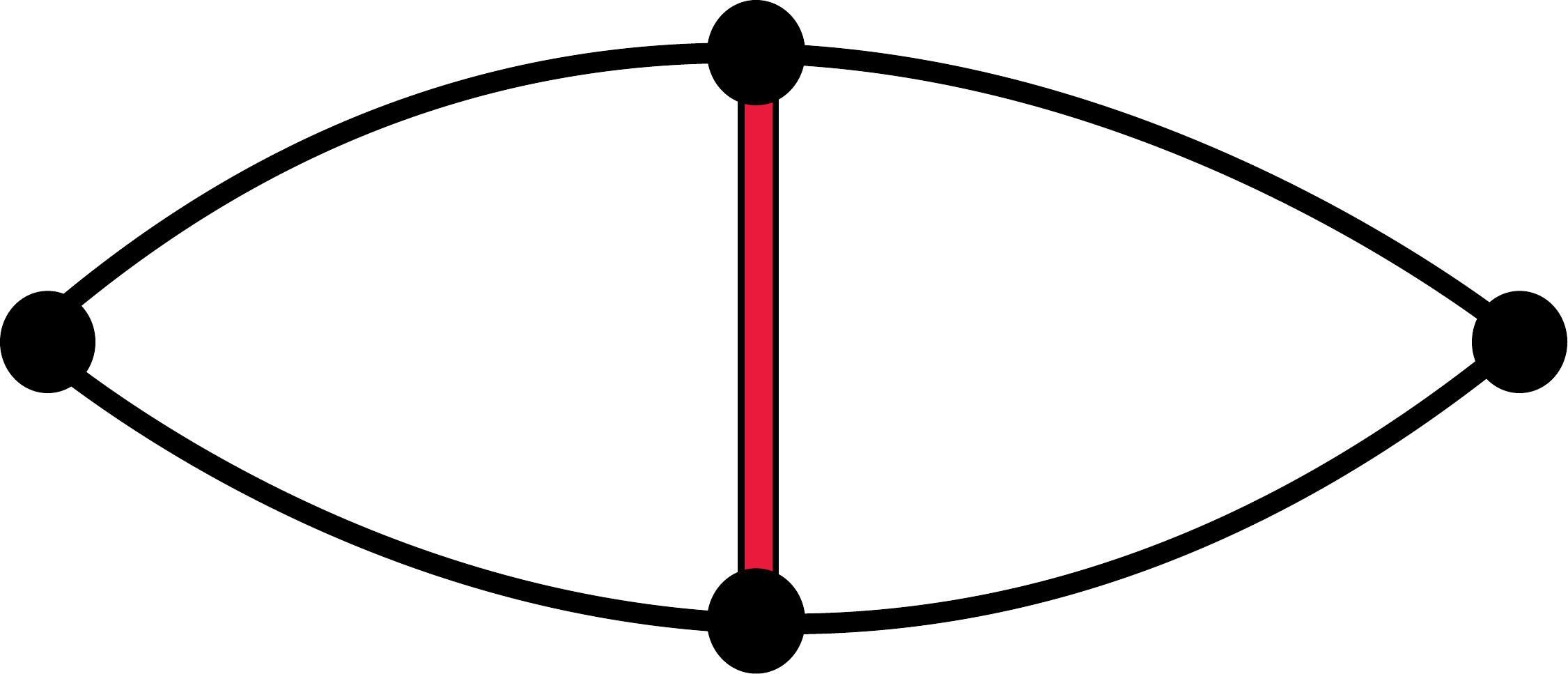}=\sqrt{2}.
\end{align}

\begin{figure}[H]
    \begin{tabular}{cc}
      \begin{minipage}[t]{0.45\hsize}
         \centering
  \includegraphics[width=4cm]{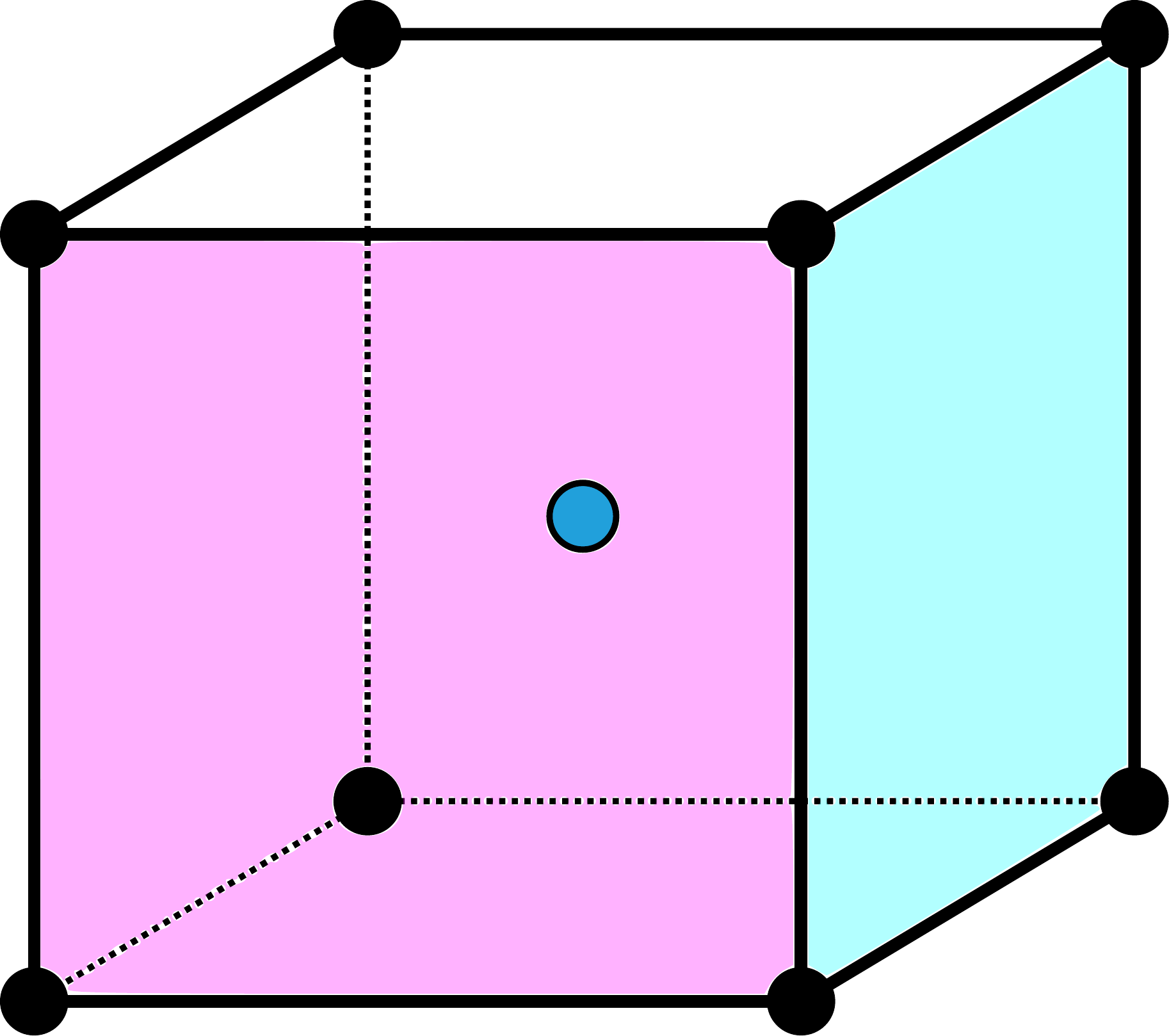}
  \caption{
  An active cube. A black line represents  an active link. 
  A blue dot represents the inactive link crossing the center of the active cube.}
  \label{fig:cube22}
      \end{minipage} &
      \begin{minipage}[t]{0.45\hsize}
        \centering
  \includegraphics[width=4cm]{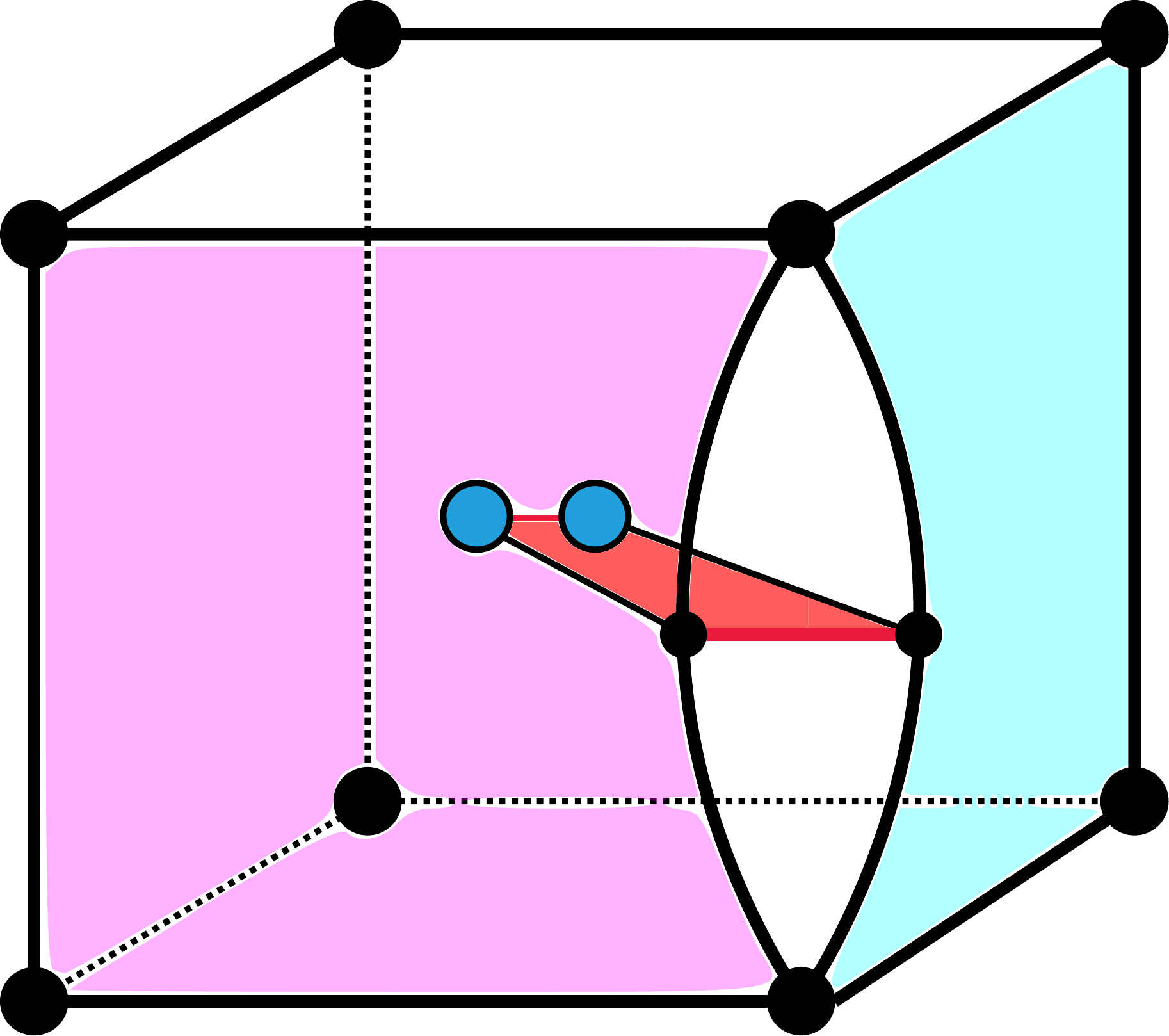}
  \caption{
  An active cube with a building block of a $\Zb_2$ symmetry defect. 
  The inactive link and an active link is doubled by the $\Zb_2$ symmetry defect. 
  Two plaquettes that share the active link before doubling contain the doubled active links, one each.}
  \label{fig:cubeZ22}
      \end{minipage}
    \end{tabular}
  \end{figure}

It is not completely obvious how to assign the Boltzmann weight for a plaquette (or a 16-cell) including doubled active links.  It is assigned by the following rules.
There is a pair of doubled inactive links in each building block of a $\Zb_2$ symmetry defect.
In our setup, an inactive link crosses the center of a cube formed by six active plaquettes, which we call an ``active cube.''
Moreover, the building block of the $\Zb_2$ symmetry defect specifies the midpoint of an active link contained in this active cube. 
Let us choose such a building block and explain how to assign the Boltzmann weights.  There are two plaquettes in this active cube sharing this active link.  
Then the Boltzmann weight of one of these two plaquettes is calculated by using one of the doubled active links, and that of the other plaquette is calculated by the other active link (see Figures \ref{fig:cube22}, \ref{fig:cubeZ22}).
It could be a bit confusing when two plaquettes in an active cube share a pair of doubled active links, although the triangle formed by the center of these doubled active links and the inactive link at the center of this active cube is not included in a $\Zb_2$ symmetry defect. 
In this case, the Boltzmann weights of both of these two plaquettes are calculated by one of these doubled active links.  These rules determine the Boltzmann weights for all plaquettes.

Now, we show that these $\Zb_2$ 1-form symmetry defects are topological. We focus on a 16-cell as in the case of the duality defects.  We denote the set of active links in this 16-cell as $M=\{m=1,2,3,4\}$ and the set of the inactive links as $\tilde N=\{\tilde n=\tilde1,\tilde2,\tilde3,\tilde4\}$.
Let us choose two neighboring active links $m=1,2$ from the set $M$.
Moreover, we focus the set of triangles
\begin{align}
V=\{(m,\tilde n)|m=1,2,\ \tilde n=\tilde1,\tilde2,\tilde3,\tilde4\}.\label{eq:Z2tetora}
\end{align}
These triangles form an octahedron.
We consider a $\Zb_2$ 1-form symmetry defect consisting of $H\subset V$ in the octahedron; this configuration is denoted by \confB.
We compare it to the deformed configuration \confC in which the defect consists $\tilde H:=V\setminus H$ in the octahedron and the configuration away from the octahedron is not changed.
In order to write down the commutation relations between these two configurations of defects, we also define $M'=\{m=1,2\}\subset M,\ F_b:=\{m|(m,\tilde n)\in \tilde H\}$ and $F_c:=\{m|(m,\tilde n)\in H\}$.  The link variables of the active links $3,4$ are denoted by $a_3,a_4$, respectively.
The link variables of the links $m=1,2$ are denoted by $b_m$ and $c_m$ if the link $m$ is doubled.  If $m=1,2$ is not doubled, its link variable is denoted by $b_m$ in the configuration \confB and $c_m$ in the configuration \confC.
Then our $\Zb_2$ symmetry defects turn out to satisfy the commutation relation
\begin{align}
\label{eq;Zcom relation}
\sum_{M'\setminus F_b}W(b_{1},b_{2},a_3,a_4)\prod_{(m,\tilde n)\in H}Z_2(b_{m},c_{m})
=\sum_{M'\setminus F_c}W(c_{1},c_{2},a_3,a_4)\prod_{(m,\tilde n)\in \tilde H}Z_2(b_{m},c_{m}).
\end{align}
Here, taking the weight of $z$ into account, the weights of sites and links on both sides are the same and cancel to each other. So we omitted them in Eq.~\eqref{eq;Zcom relation}.
The summations in Eq.~\eqref{eq;Zcom relation} are defined by 
\begin{align}
\sum_{M'\setminus F_b}:=&\prod_{m\in M'\setminus F_b}\sum_{b_m=0,1},\\
\sum_{M'\setminus F_c}:=&\prod_{m\in M'\setminus F_c}\sum_{c_m=0,1}.
\end{align}
The commutation relations \eqref{eq;Zcom relation} imply our $\Zb_2$ symmetry defects are topological.

The commutation relations \eqref{eq;Zcom relation} also imply our $\Zb_2$ symmetry defects are invertible.  
For example, $\Zb_2$ symmetry defect placed on the octahedron has the same weight as the empty configuration
\begin{align}
\sum_{b_{1},b_{2}=0}^{1}W(b_{1},b_{2},a_3,a_4)\prod_{(m,\tilde n)\in V}Z_2(b_{m},c_{m})=W(c_{1},c_{2},a_3,a_4).
\end{align}
We conclude that our $\Zb_2$ symmetry defects are actually the symmetry defects associated to the 1-form $\Zb_2$ center symmetry.

\subsection{Defect junctions}
\label{sec:defect junction}
We can consider a configuration of defects where two or more kinds of defects meet and make junctions.  In this subsection, we discuss such junctions and their weights.

\begin{figure}[H]
  \centering
  \includegraphics[width=5cm]{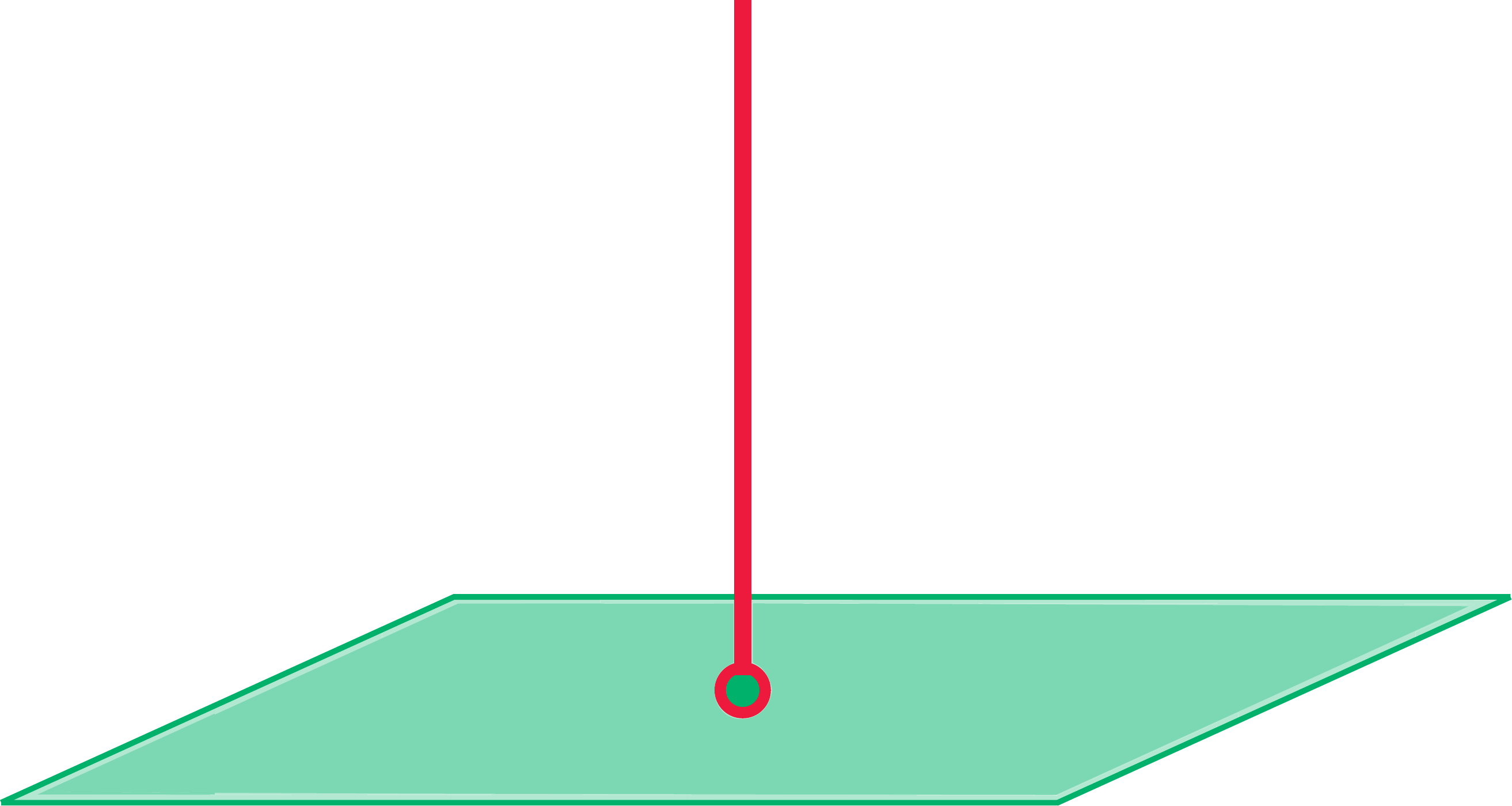}
  \caption{A schematic illustration of junctions. The red line represents a $\Zb_2$ symmetry defect, and the green surface represents a duality defect. The intersection is actually 1-dimensional.}
  \label{fig:defectjunction}
\end{figure}

In our system, a junction occurs when a $\Zb_2$ 1-form center symmetry defect ends on a duality defect as in Figure \ref{fig:defectjunction}. 
Junctions are located on 1-dimensional lines.

\begin{figure}[H]
    \begin{tabular}{cc}
      \begin{minipage}[t]{0.45\hsize}
         \centering
        
  \includegraphics[width=3cm]{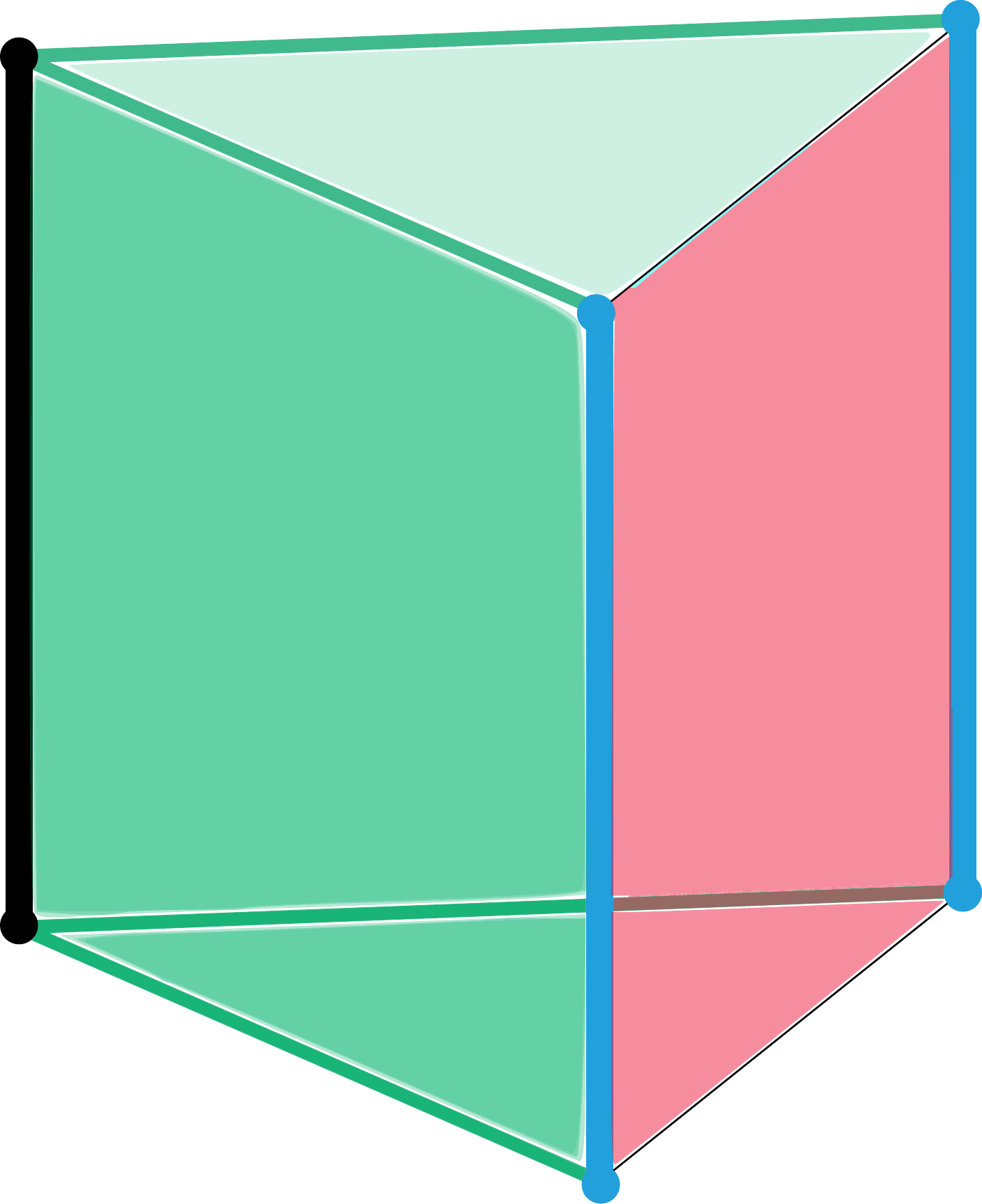}
  \caption{A schematic illustration of a junction sharing an inactive link. Its weight is denoted as $J(a)$.}
 	\label{fig:inactive_defectjuncyion}
      \end{minipage} &
      \begin{minipage}[t]{0.45\hsize}
        \centering
        
  \includegraphics[width=4cm]{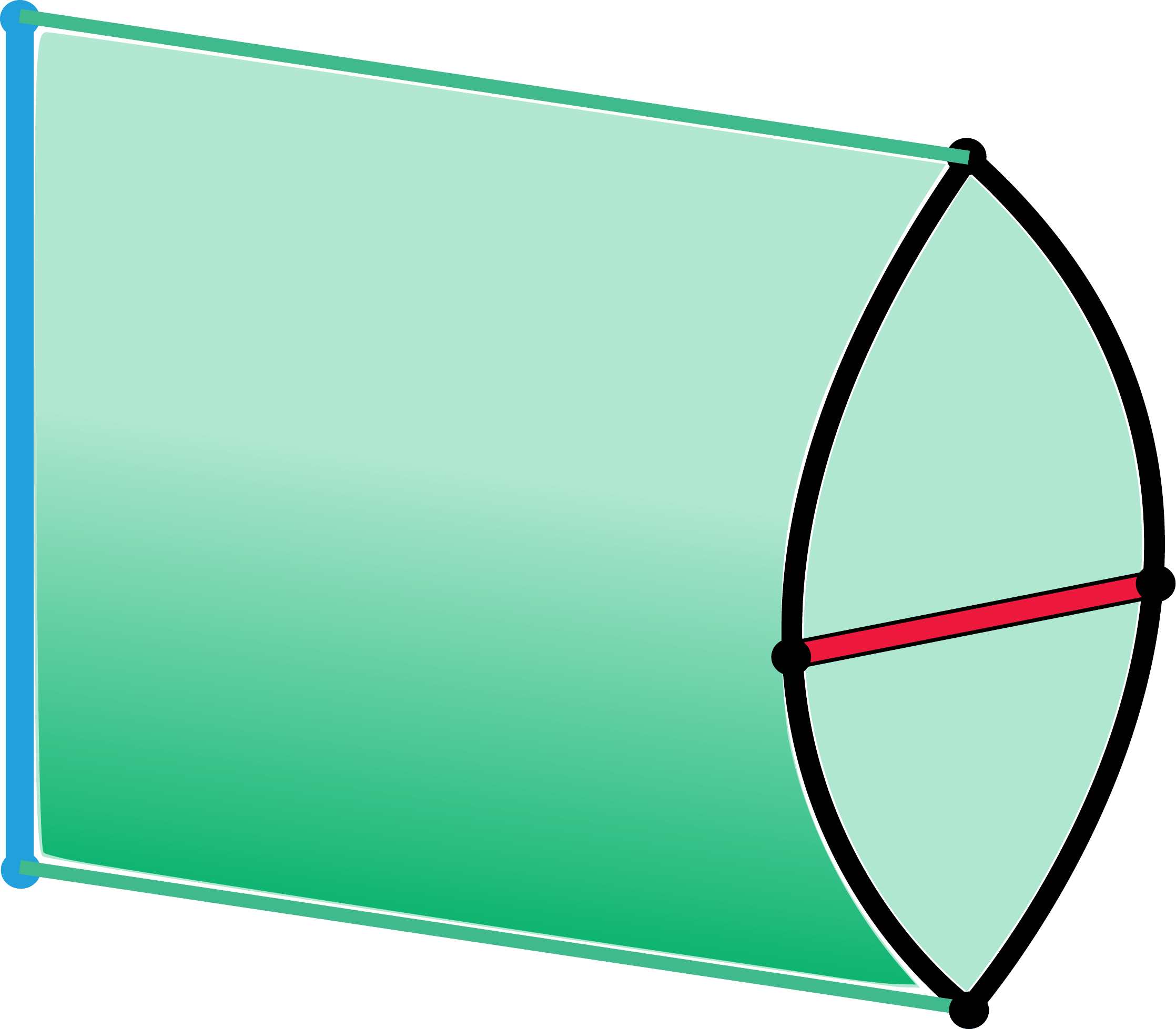}
  \caption{A schematic illustration of a junction sharing an active link. Its weight is denoted as $\tilde{J}(b,c)$.}
	\label{fig:active_defectjuncyion}
      \end{minipage}
    \end{tabular}
  \end{figure}

As discussed in Sec.~\ref{sec:KWdefect}, duality defects are defined by doubling tetrahedrons and links on them. 
A junction occurs when a doubled link of the duality defect and the $\Zb_2$ symmetry defect have an intersection. 
There are two types of junctions, depending on whether the common link is an active link or an inactive link. Each junction has the form shown in Figure \ref{fig:inactive_defectjuncyion} and \ref{fig:active_defectjuncyion}.
The weight of a junction when sharing an inactive link is denoted as $J(a)$ with the link variable $a$ of the active link corresponding to the common inactive link. 
Similarly, if the common link is an active link, the weight of the arising junction is denoted by $\tilde{J}(b,c)$, where $b,c$ are the link variables of the common links doubled by the $\Zb_2$ symmetry defect.
Since the total weight is non-zero only when $b=1-c$ due to the $\Zb_2$ symmetry defect, we can use a function $\tilde{J}(a)$ which depends only on $a$ and express the weight of this junction as $\tilde{J}(b,c)=\tilde{J}(b)\sigma^x_{b,c}$. 

\begin{figure}[htbp]
  \centering
  \includegraphics[width=7cm]{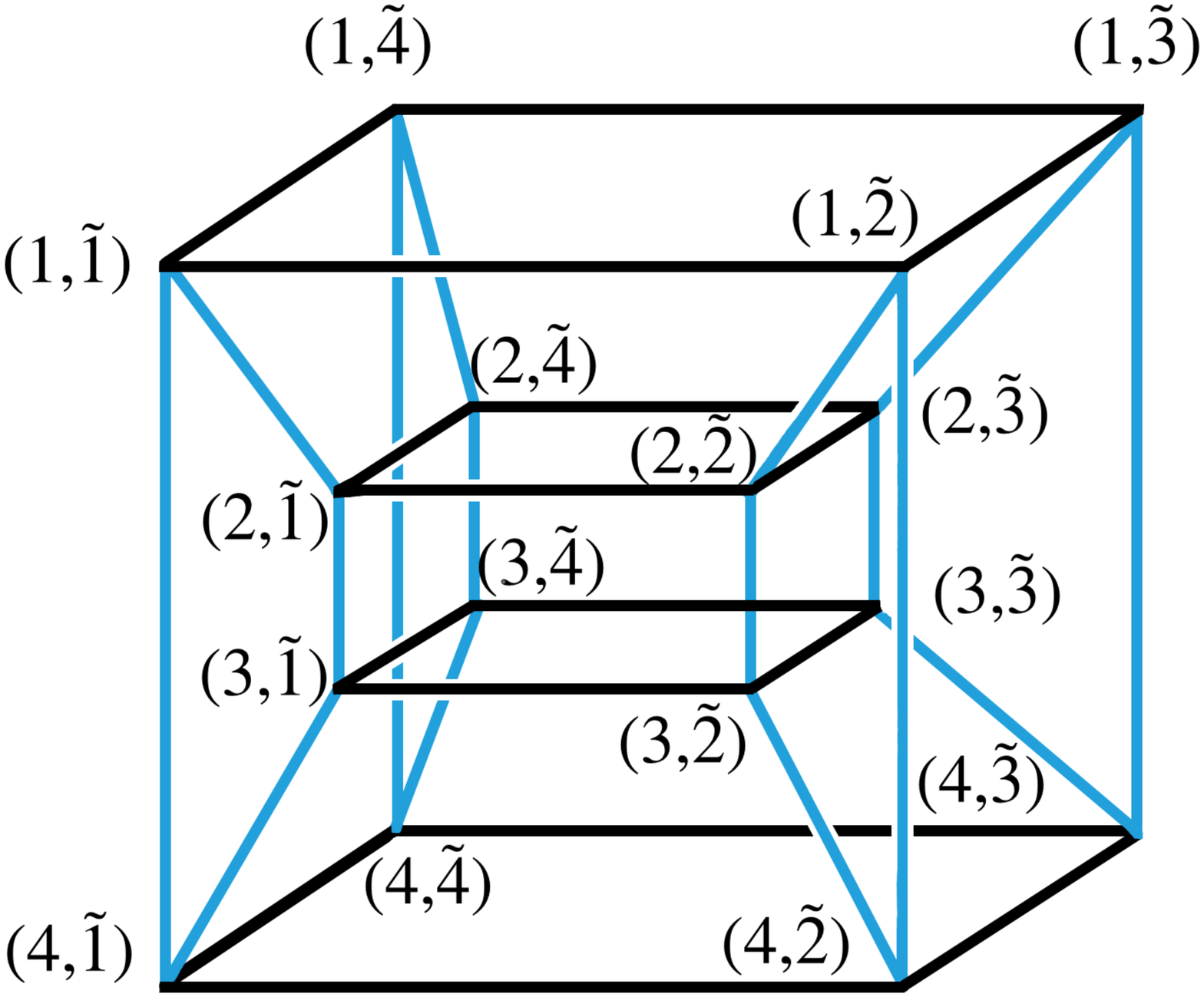}
  \caption{A configuration of tetrahedrons. Each vertex represents a tetrahedron. Two tetrahedrons connected by the black line share an active link, and two tetrahedrons connected by the blue line share an inactive link. The pairs of numbers assigned to the vertices specify the labels of the active and inactive links that each tetrahedron contains.}
  \label{fig:16cells2}
\end{figure}

These junction weights are determined so that the following junction commutation relations are satisfied.
In order to explain junction commutation relations concretely, we introduce a diagram as shown in Figure \ref{fig:16cells2}.
In this diagram, each vertex represents a tetrahedron. Two tetrahedrons connected by a black line share an active link, and two tetrahedrons connected by a blue line share an inactive link.

\begin{figure}[htbp]
  \centering
  \includegraphics[width=15cm]{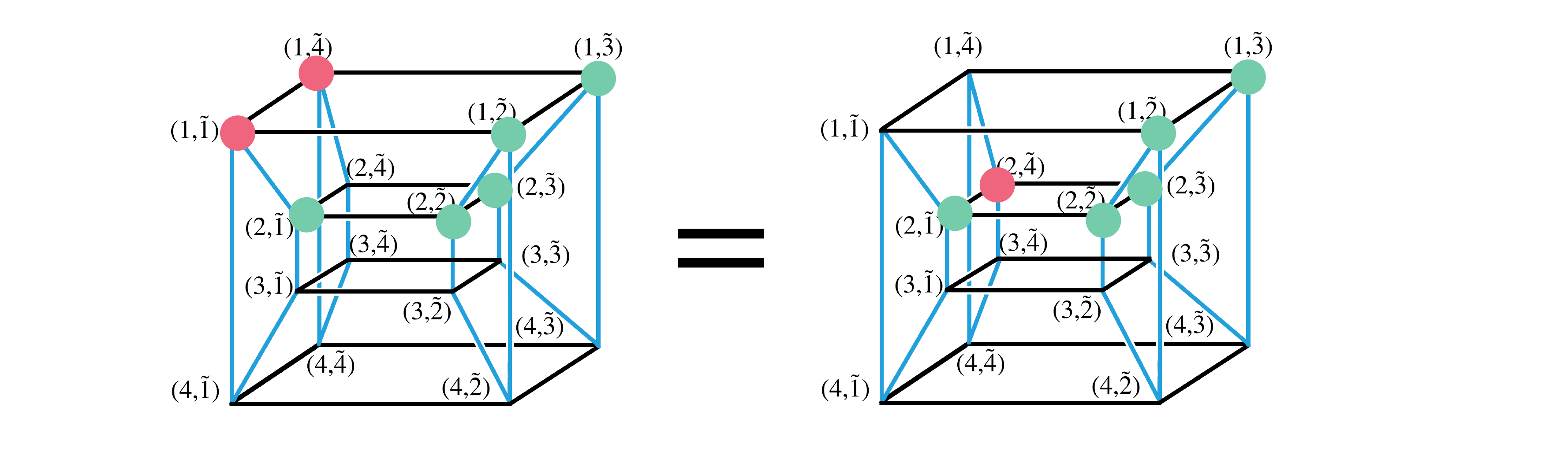}
  \caption{The configuration of Eq.~\eqref{eq:inactivejunc_rel}. 
The green dots represent the tetrahedrons on which the duality defect is located.
The red dots represent the tetrahedrons on which the $\Zb_2$ symmetry defect is located.  
Here, placing a $\Zb_2$ symmetry defect on a tetrahedron means placing a $\Zb_2$ symmetry defect on a triangle connecting the center of the active link with the inactive link in the tetrahedron.}
 \label{fig:juncrel1}
  \end{figure}

\begin{figure}[htbp]
  \centering
  \includegraphics[width=7cm]{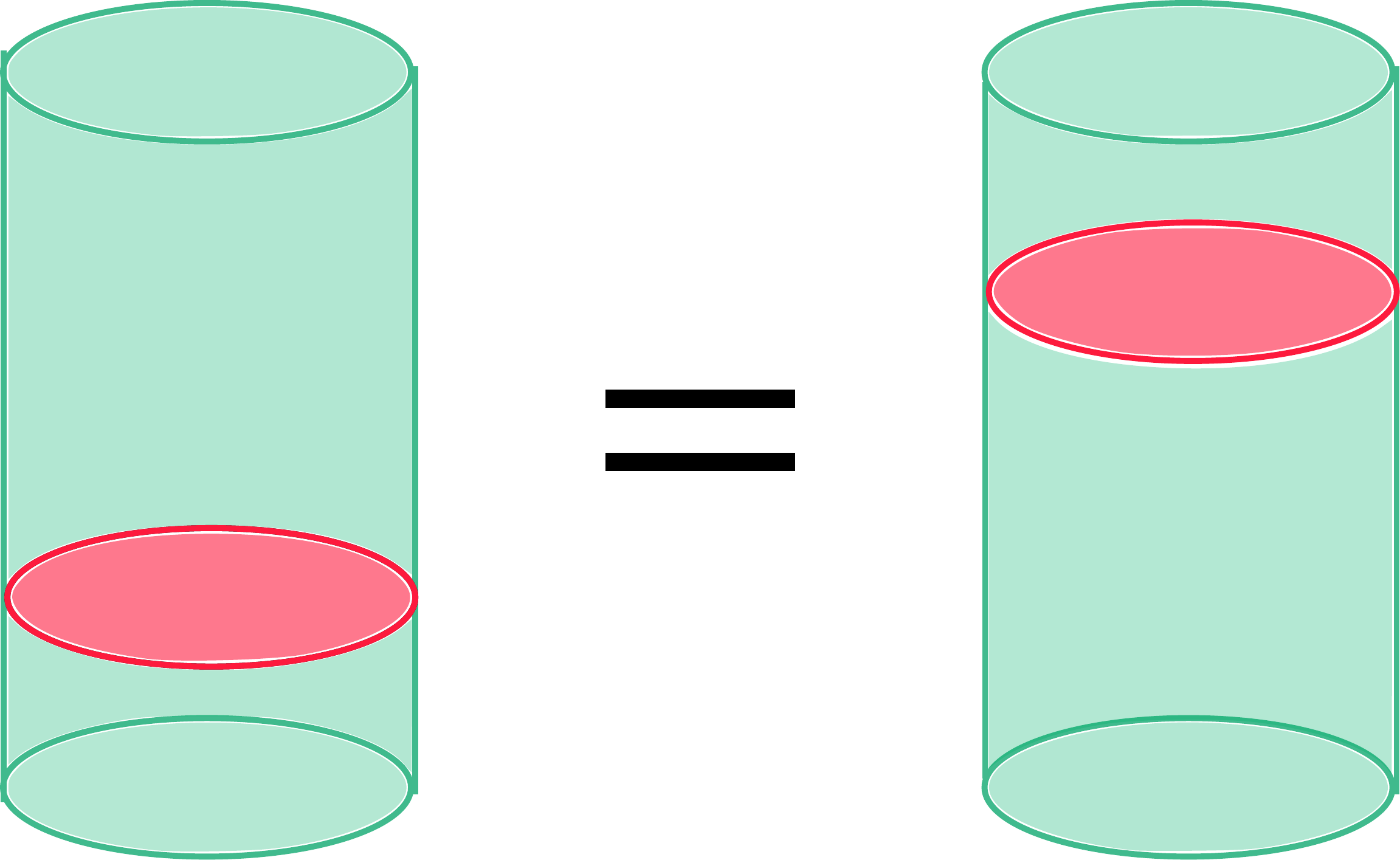}
  \caption{A schematic illustration of junction commutation relations. The green surface represents a duality defect and the red surface represents a $\Zb_2$ symmetry defect. 
The junction commutation relation requires that the $\Zb_2$ defect can be deformed along with the duality defect.  Junctions are located on the intersection of the two kinds of defects.}
   \label{fig:juncrel_illust}
\end{figure}

First, we impose a junction commutation relation of the $J$ junction weights and the $\tilde{J}$ junction weights.
We consider two configurations of defects shown in Figure \ref{fig:juncrel1} and require their weights are equal to each other.
This commutation relation is a requirement that the $\Zb_2$ symmetry defect can be continuously deformed along with the duality defect as schematically depicted in Figure \ref{fig:juncrel_illust}.
This commutation relation can be written as
\begin{align}
	&W(1-a_1,a_2,a_3,a_4) D(1-a_1,\tilde{a}_{\tilde{2}}) D(1-a_1,\tilde{a}_{\tilde{3}})  D(a_2,\tilde{a}_{\tilde{1}}) D(a_2,\tilde{a}_{\tilde{2}}) D(a_2,\tilde{a}_{\tilde{3}})J(\tilde{a}_{\tilde{1}})\tilde{J}(1-a_1)\nonumber\\
	&=W(a_1,1-a_2,a_3,a_4) D(a_1,\tilde{a}_{\tilde{2}})  D(a_1,\tilde{a}_{\tilde{3}}) D(1-a_2,\tilde{a}_{\tilde{1}}) D(1-a_2,\tilde{a}_{\tilde{2}}) D(1-a_2,\tilde{a}_{\tilde{3}})\tilde{J}(1-a_2).
  \label{eq:inactivejunc_rel}
\end{align}
Here, the constant weights are omitted since they are common to both sides. The summation caused by the $\Zb_2$ symmetry defect is eliminated.

\begin{figure}[htbp]
  \centering
  \includegraphics[width=15cm]{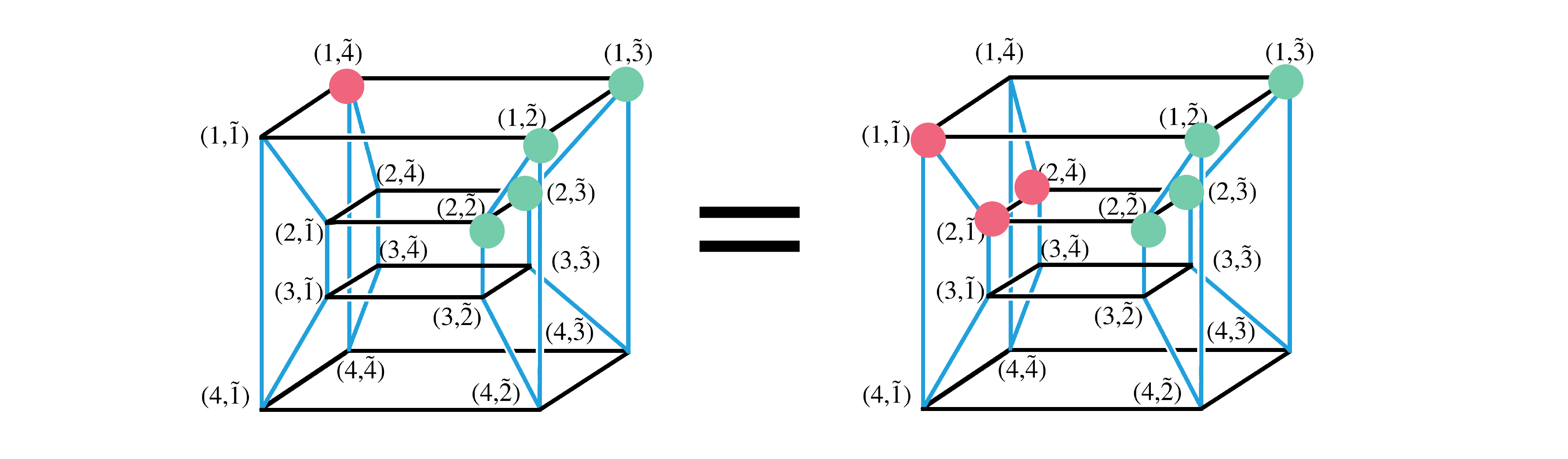}
  \caption{The configuration of Eq.~\eqref{eq:activejunc_rel}.}
\label{fig:juncrel2}
\end{figure}

Next, we determine the $\tilde{J}(a,b)$ junction weight from junction commutation relations. We consider two configurations of defects shown in Figure \ref{fig:juncrel2}, and require that their weights are equal to each other.
This junction commutation relation is also a requirement that the $\Zb_2$ symmetry defect can be continuously deformed along with the duality defect.
This equation reads
\begin{align}
	&W(1-a_1,a_2,a_3,a_4) D(1-a_1,\tilde{a}_{\tilde{2}}) D(1-a_1,\tilde{a}_{\tilde{3}})D(a_2,\tilde{a}_{\tilde{2}})D(a_2,\tilde{a}_{\tilde{3}}) \tilde{J}(1-a_1)\nonumber\\
	&=W(a_1,1-a_2,a_3,a_4) D(a_1,\tilde{a}_{\tilde{2}}) D(a_1,\tilde{a}_{\tilde{3}}) D(1-a_2,\tilde{a}_{\tilde{2}})D(1-a_2,\tilde{a}_{\tilde{3}}) \tilde{J}(a_1)\tilde{J}(1-a_2).
	\label{eq:activejunc_rel}
\end{align}
Here, the constant weights are omitted since they are common to both sides, and the summation caused by the $\Zb_2$ symmetry defect has been already eliminated.

We can solve the junction commutation relations of Eq.~\eqref{eq:inactivejunc_rel} and Eq.~\eqref{eq:activejunc_rel}, and find a unique solution for $J(a), \tilde J (b,c)$. It is written as
\begin{align}
	J(a)&=\includegraphics[width=1.3cm,align=c]{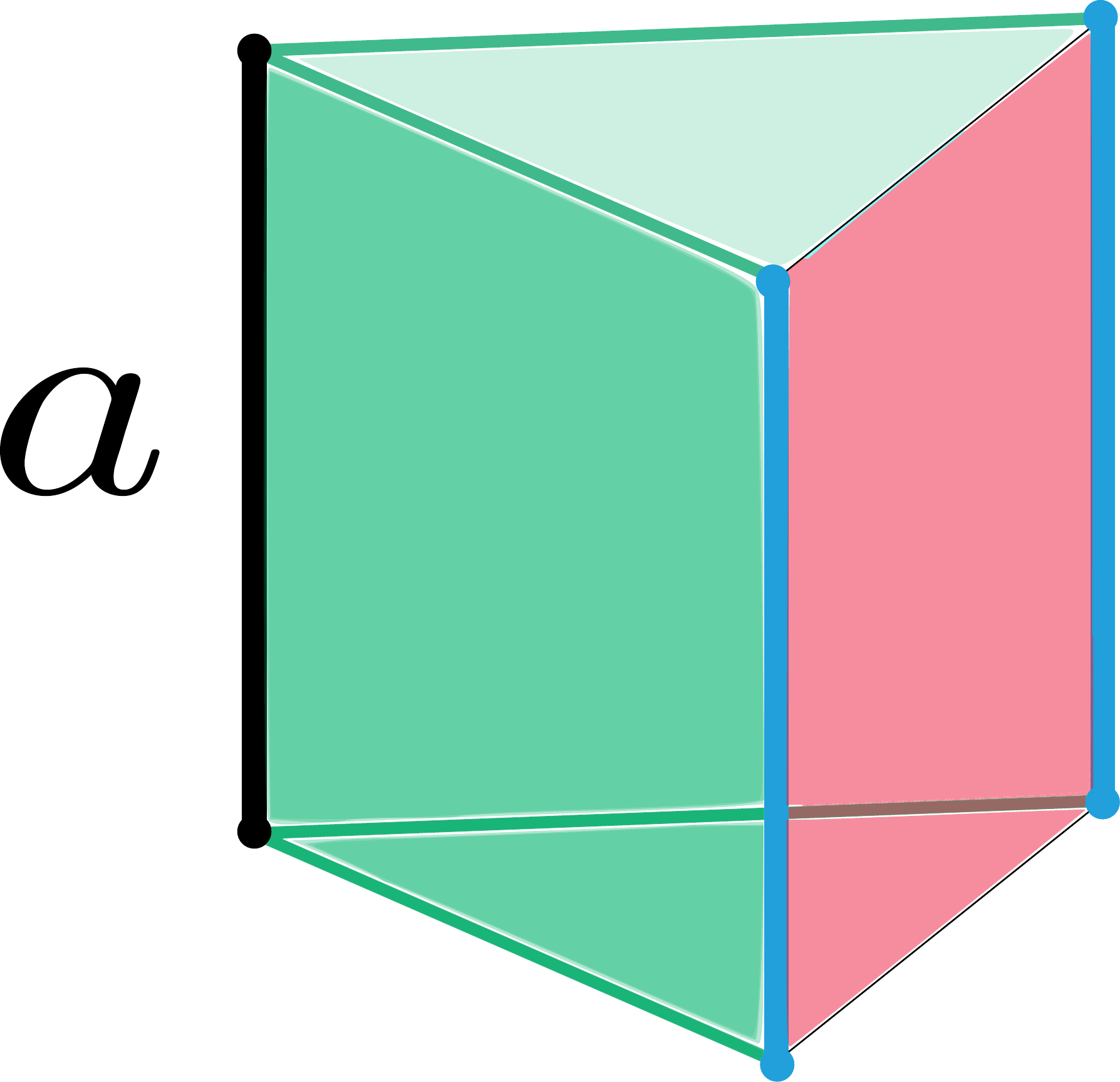}=(-1)^a,\label{J_solution}\\
	\tilde{J}(b,c)&=\includegraphics[width=1.3cm,align=c]{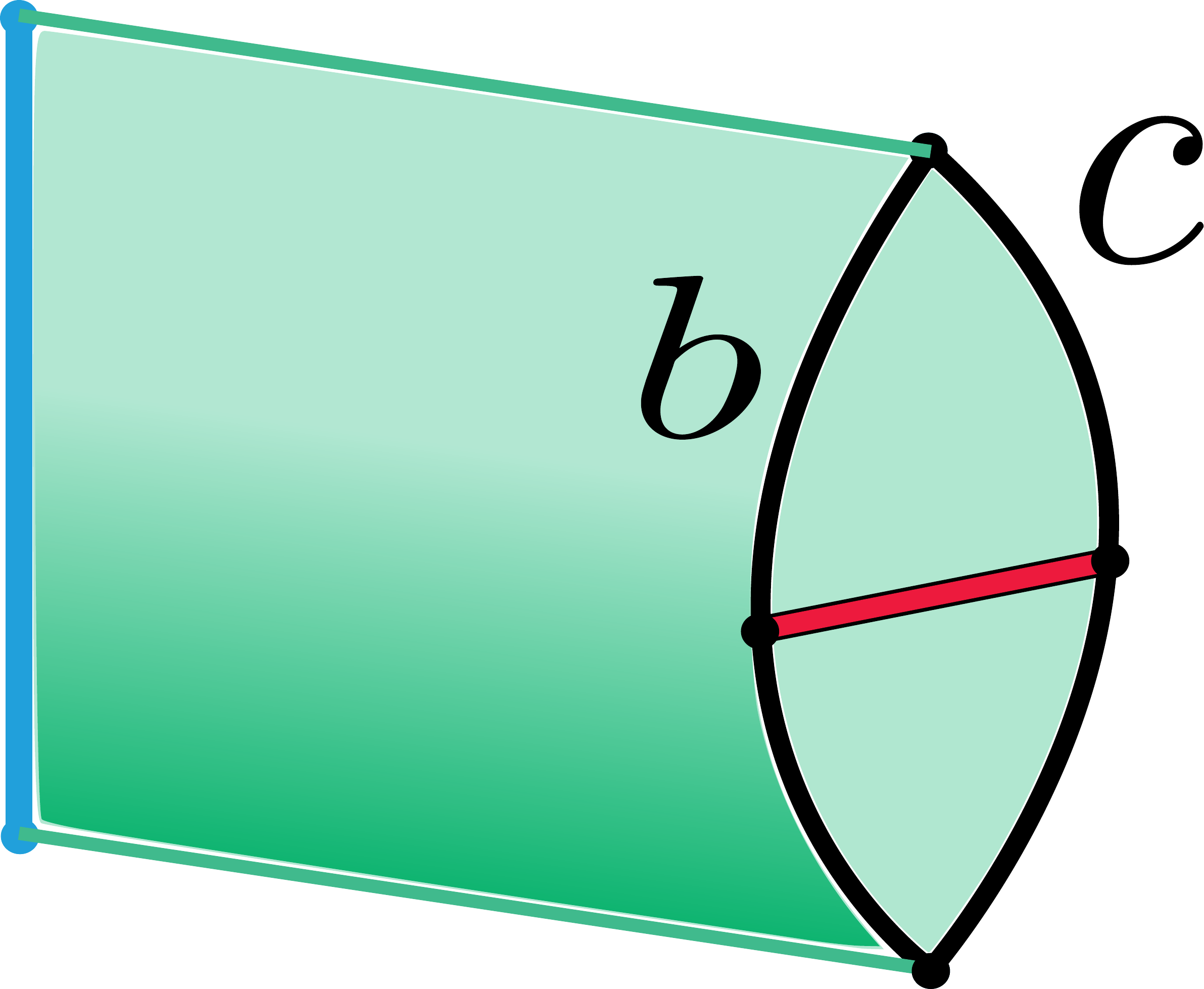}=\sigma^x_{b,c}.\label{Jtilde_solution}
\end{align}

There are general junction commutation relations including Eqs.~\eqref{eq:inactivejunc_rel}, \eqref{eq:activejunc_rel}.
We explain these commutation relations using the notation of Sec.~\ref{sec:KWdefect}.
Let us suppose duality defects are placed on $I\subset U$.
We choose $i,j\in M=\{1,2,3,4\}\ (i\neq j)$ and consider two configurations of $\Zb_2$ symmetry defects on $F=\{i,j\}\times \tilde{N},\ \tilde{N}:=\{\tilde{1},\tilde{2},\tilde{3},\tilde{4}\}$.
We define $I_F=\{(m,\tilde{n})|(m,\tilde{n})\in I\ \mbox{and}\ m\in\{i,j\}\}$ as the set of building blocks of the duality defect whose active links are $i$ or $j$. We also define $\bar{F}=F\setminus I_F$. 
We require that the weight of the configuration with a $\Zb_2$ defect located on $H_F\subset \bar{F}$ is equal to that on $\bar{F}\setminus H_F$.
These are non-trivial relations for junction weights.
The weights \eqref{J_solution}, \eqref{Jtilde_solution} satisfy all these commutation relations. 

Let us explain the gauge invariance of our junctions.  The weight \eqref{Jtilde_solution} is gauge invariant by itself.
This is because it transforms as $\tilde{J}(b,c) \to \tilde{J}(1-b,1-c)=\tilde{J}(b,c)$ by the gauge transformation at an active site contained in this building block.
On the other hand, $J(a)$ in Eq.~\eqref{J_solution} is not gauge invariant by itself, because it transforms as $J(a)\to J(1-a)=-J(a)$ by the gauge transformation at an active site contained in this building block of junctions.
However a whole junction is gauge invariant as follows. 
Let us focus on an active site $S$ on a $J$ junction.  
Since this junction is a line without boundaries, 
there are exactly two building blocks of junctions which include $S$.
Let $a,b$ the link variables of the two active links in these two building blocks. 
By the gauge transformation at $S$, the weights of these two building blocks transform as $J(a)J(b)\to J(1-a)J(1-b)=J(a)J(b)$.
The other part of the junction is trivially invariant by this gauge transformation.
Therefore we can conclude that our junctions are gauge invariant.

\subsection{Crossing relations and expectation values}
\label{sec:commutationrelations}
In this subsection, we discuss various crossing relations using the defects and junctions described so far.
For example, we derive the crossing relations between two duality defect configurations whose topologies are different.
We also derive some crossing relations in which both duality defects and $\Zb_2$ symmetry defects appear.

\begin{figure}[htbp]
  \centering
  \includegraphics[width=5cm]{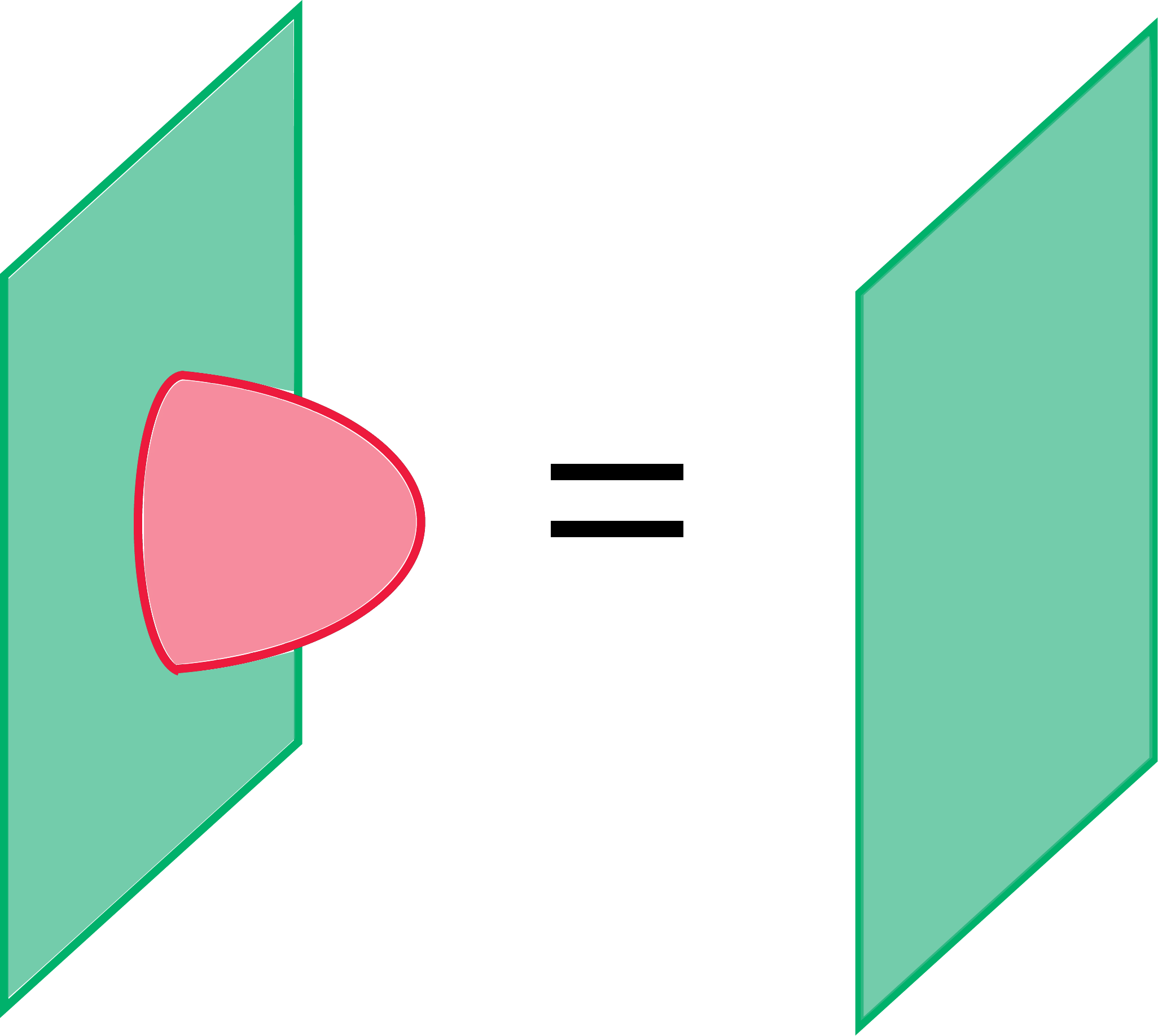}
  \caption{A schematic illustration of the crossing relations that a $\Zb_2$ symmetry defects with a boundary on the duality defect can be removed if the boundary is homologically trivial on the duality defect.}
  \label{fig:commutation_relation1}
\end{figure}

\begin{figure}[htbp]
  \centering
  \includegraphics[width=5cm]{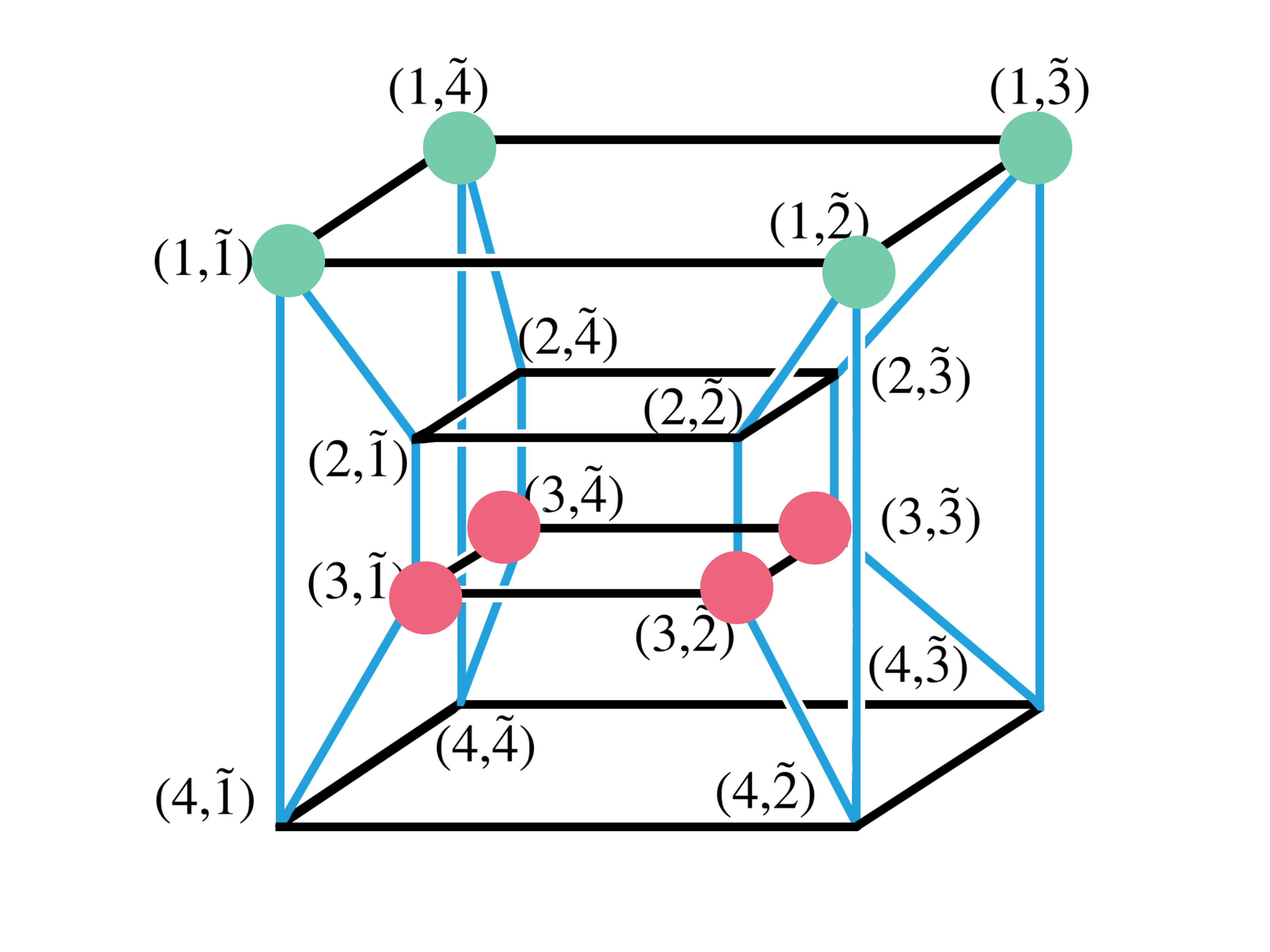}
  \caption{ The configuration of the left-hand side of Eq.~\eqref{eq:commutation relation1}. }
  \label{fig:commutation_relation1_place}
\end{figure}

An example of such relations is that $\Zb_2$ symmetry defects with a boundary on the duality defect can be removed if the boundary is homologically trivial on the duality defect.  It is schematically illustrated in Figure \ref{fig:commutation_relation1}.
One concrete configuration in the lattice is shown in Figure \ref{fig:commutation_relation1_place}.
We can verify the following equation that represents Figure \ref{fig:commutation_relation1_place}.
\begin{align}
	\sum_{a_1,b_3=0,1}J(\tilde{a}_{\tilde{1}})J(\tilde{a}_{\tilde{2}})J(\tilde{a}_{\tilde{3}})J(\tilde{a}_{\tilde{4}})W(a_1,a_2,b_3,a_4)Z_2(b_3,c_3)^4D(a_1,\tilde{a}_{\tilde{1}})D(a_1,\tilde{a}_{\tilde{2}})D(a_1,\tilde{a}_{\tilde{3}})D(a_1,\tilde{a}_{\tilde{4}})\nonumber\\
	=\sum_{a_1=0,1}W(a_1,a_2,c_3,a_4)D(a_1,\tilde{a}_{\tilde{1}})D(a_1,\tilde{a}_{\tilde{2}})D(a_1,\tilde{a}_{\tilde{3}})D(a_1,\tilde{a}_{\tilde{4}}).\label{eq:commutation relation1}
\end{align}
Here, we omit the constant weights such as link weights because these weights are common to both sides.


\begin{figure}[htbp]
  \centering
  \includegraphics[width=10cm]{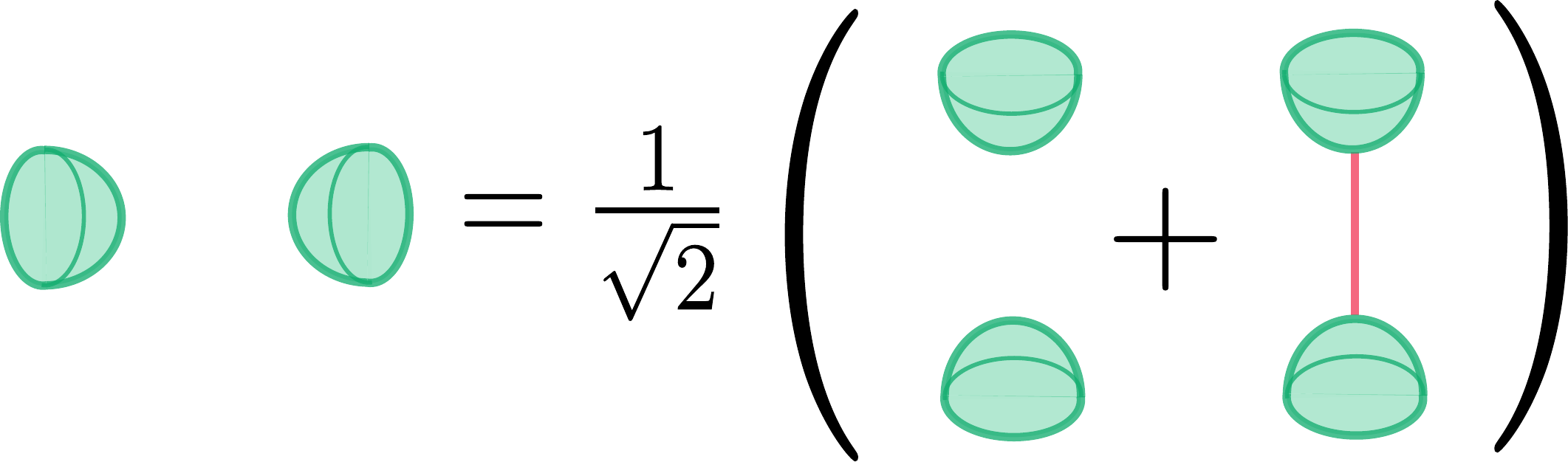}
  \caption{A schematic illustration of solid torus equations. 
The left-hand side represents a duality defect configuration on one solid torus $V_1$. 
The first term of the right-hand side represents a duality defect configuration on one solid torus $V_2$ and the second term represents a configuration of the duality defect on a solid torus $V_2$ with a $\Zb_2$ symmetry defect on $D^2$ whose boundary is a non-trivial cycle on the duality defect.
This is an example in which the crossing relation is not closed only within duality defects.
}
  \label{fig:commutation_relation2_sch}
\end{figure}

\begin{figure}[htbp]
  \centering
  \includegraphics[width=13cm]{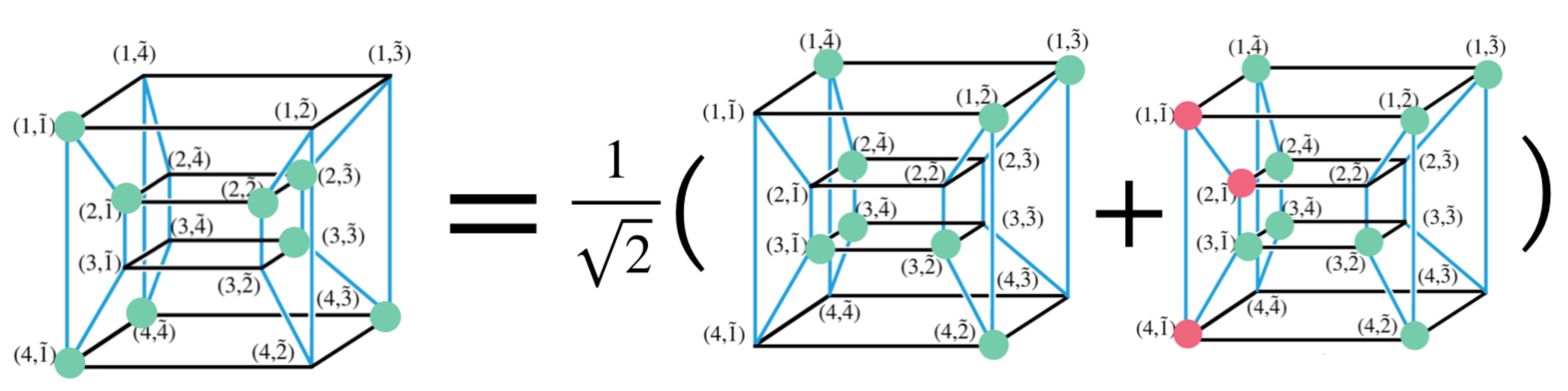}
  \caption{The configuration of Eq.~\eqref{eq:solid torus rel.}. The left-hand side is the duality defect configuration on $V_1$. The first term on the right-hand side is the duality defect configuration on $V_2$, and the second term on the right-hand side is the duality defect configuration on $V_2$ with a $\Zb_2$ symmetry defect.}
  \label{fig:commutation_relation2_place}
\end{figure}

Another important crossing relation is the equation for the duality defect on 3-dimensional solid torus.
We consider the decomposition of a 16-cell into two solid tori. 
We call these solid tori $V_1$ and $V_2$.
A solid torus is not simply connected and thus the topology of the duality defects changes by the deformation as explained in Sec.~\ref{sec:KWdefect}.
Therefore, the defect commutation relations need not be satisfied.
Instead, we find the solid torus crossing relations including $\Zb_2$ defects (see Figure \ref{fig:commutation_relation2_sch}); there is the relation among the duality defect on $V_1$, the duality defect on $V_2$ and the duality defect on $V_2$ with the $\Zb_2$ symmetry defect of the following.
Let $I_S$ be the set of tetrahedrons in $V_2$.
First, we choose one $\tilde{k}\in N=\{\tilde{1},\tilde{2},\tilde{3},\tilde{4}\}$, and define $I_S^{\tilde{k}}:=\{(m,\tilde{k})|(m,\tilde{k})\in I_S\}$ to be the set of tetrahedrons in $V_2$ that contain $\tilde{k}$.
Consider a duality defect on $V_2$ with a $\Zb_2$ symmetry defect on every tetrahedron on $M\times\{\tilde{k}\}\setminus I_S^{\tilde{k}},\ M=\{1,2,3,4\}$. 
Here, placing the $\Zb_2$ symmetry defect on the tetrahedron means placing the $\Zb_2$ symmetry defect in a triangle such that it connects the center of the active link with the inactive link in the tetrahedron.  
Notice that the weight of the configuration is independent of the choice $\tilde{k}$ because of the defect commutation relations described in Sec.~\ref{sec:defect junction}.
For example, if the duality defect is placed on $V_1$ as shown in the left-hand side of Figure \ref{fig:commutation_relation2_place}, the configuration of the $\Zb_2$ symmetry defect is shown in the second term on the right-hand side of Figure \ref{fig:commutation_relation2_place}.

\begin{figure}[htbp]
  \centering
  \includegraphics[width=8cm]{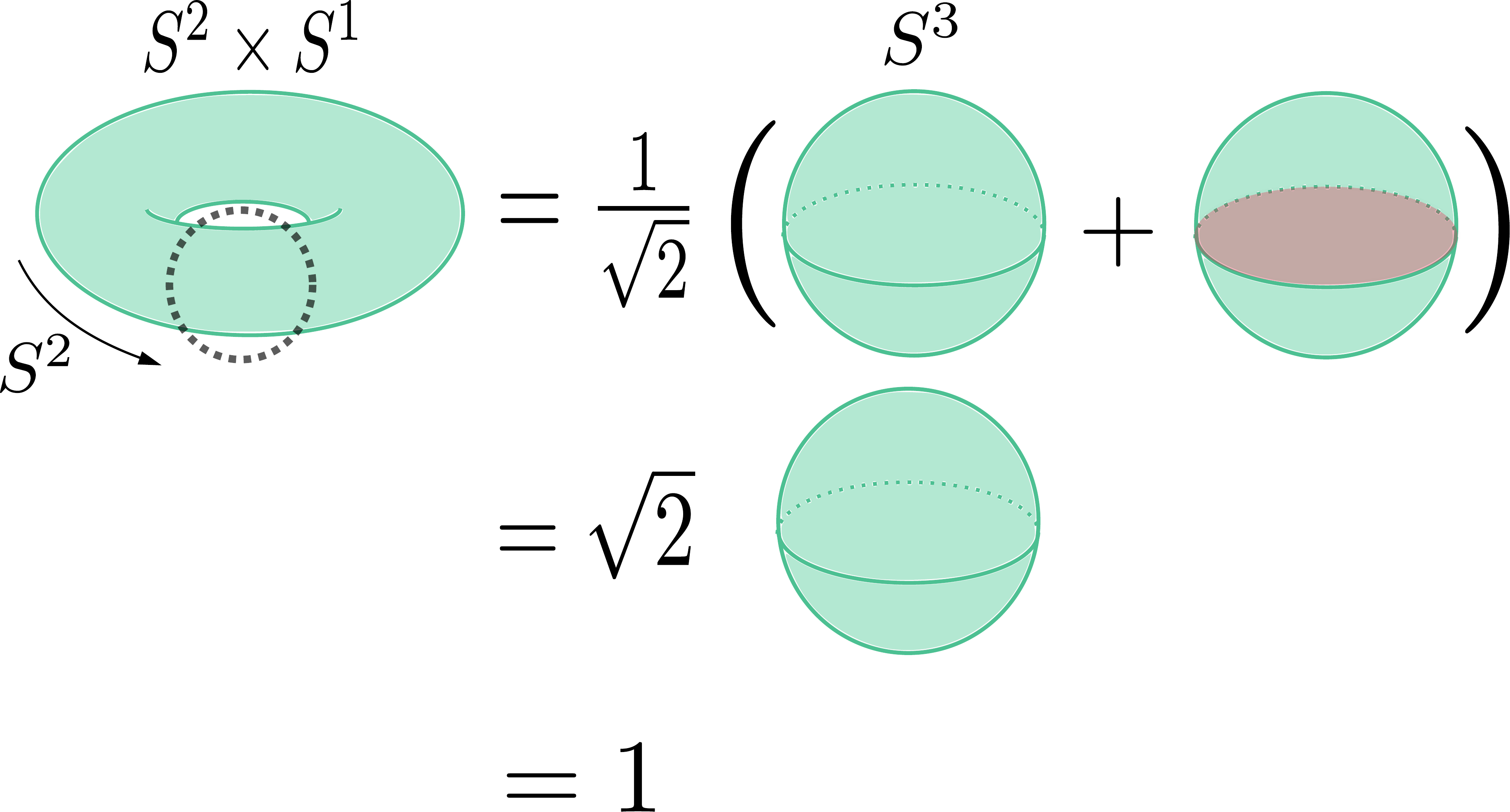}
  \caption{A calculation of an expectation value of a duality defect on $S^2\times S^1$.
In the left-hand side, we focus on the part surrounded by the dotted circle. The duality defect in this part is $D^2\times S^1$ shaped, and therefore we apply the solid torus equation here.
In the right-hand side in the first line, the first term is duality defect on $S^3$.
The second term is a duality defect on $S^3$ and a $\Zb_2$ symmetry defect on $D^2$ whose boundary is on the duality defect.
By using a crossing relation in Figure \ref{fig:commutation_relation1}, we find the second term equal to the first term.
Therefore, the expectation value of a duality defect on $S^2\times S^1$ is equal to the $\sqrt{2}$ times $S^3$ expectation value.
By using the relation in Figure \ref{fig:qdim}, finally, we find the expectation value of a duality defect on $S^2\times S^1$ is one.
}
  \label{fig:S2timesS1}
\end{figure}

\begin{figure}[htbp]
  \centering
  \includegraphics[width=9cm]{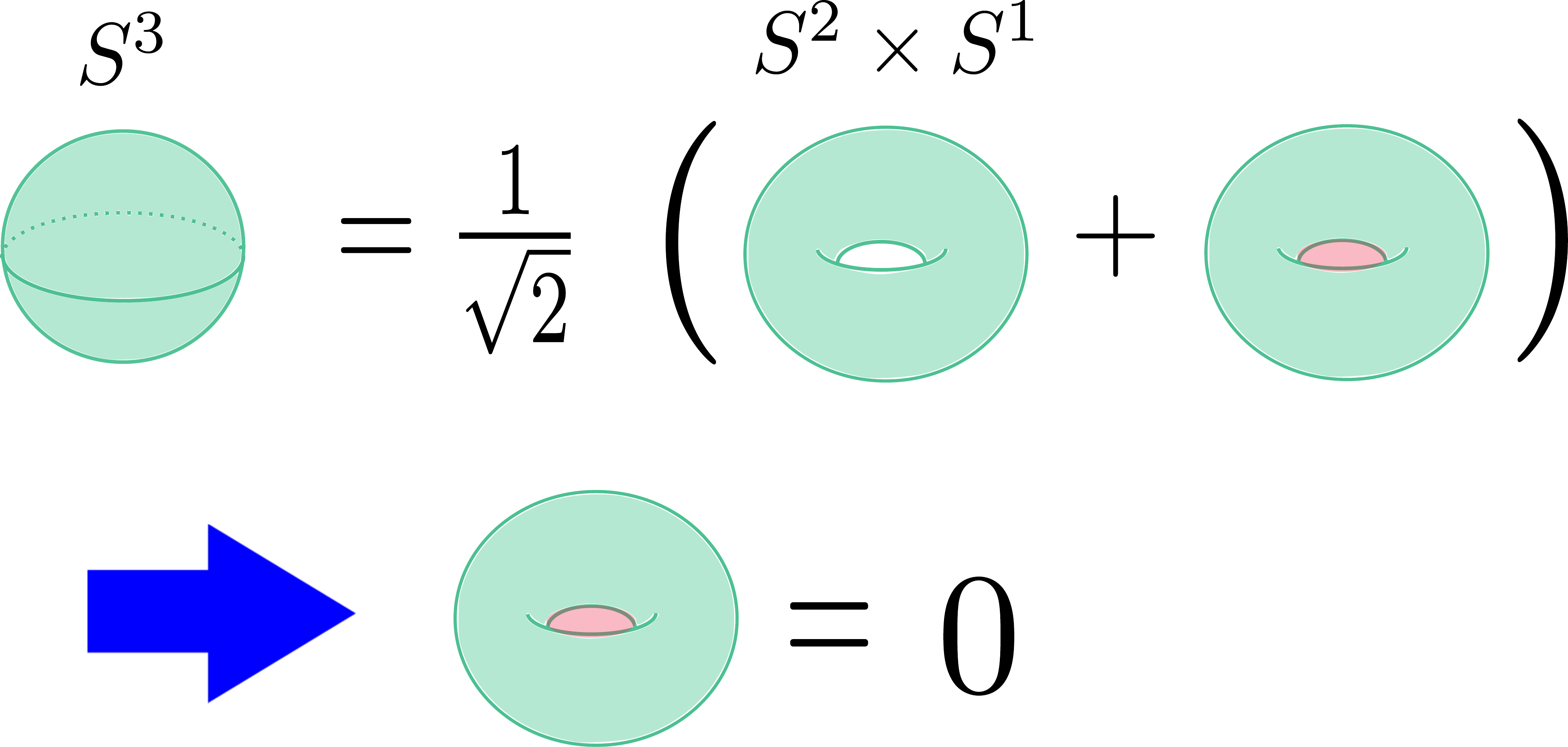}
  \caption{A calculation of the expectation value of a duality defect on $S^2\times S^1$ with a $\Zb_2$ symmetry defect on $D^2$ whose boundary is a non-trivial cycle on the duality defect.
  In the first equality, we use the solid torus equation in Figure \ref{fig:commutation_relation2_sch}.
  By using the relations in Figure \ref{fig:qdim} and \ref{fig:S2timesS1}, we find that the expectation value is zero.}
  \label{fig:S2timesS1Z2}
\end{figure}
Then, the crossing relation as shown in Figure \ref{fig:commutation_relation2_place} is satisfied. The concrete equation for this is as follows.
\begin{align}
	&s^8 l^8 W(a_1,a_2,a_3,a_4) D(a_1,\tilde{a}_{\tilde{1}}) D(a_2,\tilde{a}_{\tilde{1}}) D(a_2,\tilde{a}_{\tilde{2}})D(a_2,\tilde{a}_{\tilde{3}})D(a_3,\tilde{a}_{\tilde{3}})D(a_4,\tilde{a}_{\tilde{1}})D(a_4,\tilde{a}_{\tilde{3}})D(a_4,\tilde{a}_{\tilde{4}})\nonumber\\
	&=\frac{1}{\sqrt{2}}\Big(s^8 l^8 W(\tilde{a}_{\tilde{1}},\tilde{a}_{\tilde{2}},\tilde{a}_{\tilde{3}},\tilde{a}_{\tilde{4}})D(a_1,\tilde{a}_{\tilde{2}})D(a_1,\tilde{a}_{\tilde{3}})D(a_1,\tilde{a}_{\tilde{4}})D(a_2,\tilde{a}_{\tilde{4}})D(a_3,\tilde{a}_{\tilde{1}})D(a_3,\tilde{a}_{\tilde{2}})D(a_3,\tilde{a}_{\tilde{4}})D(a_4,\tilde{a}_{\tilde{2}})\nonumber\\
	&+\sum_{\tilde{b}_{\tilde{1}}=0,1} s^8 l^9 z W(\tilde{b}_{\tilde{1}},\tilde{a}_{\tilde{2}},\tilde{a}_{\tilde{3}},\tilde{a}_{\tilde{4}})D(a_1,\tilde{a}_{\tilde{2}})D(a_1,\tilde{a}_{\tilde{3}})D(a_1,\tilde{a}_{\tilde{4}})D(a_2,\tilde{a}_{\tilde{4}})D(a_3,\tilde{b}_{\tilde{1}})D(a_3,\tilde{a}_{\tilde{2}})\nonumber\\
	&\qq{}\qq{}\qq{}\times D(a_3,\tilde{a}_{\tilde{4}})D(a_4,\tilde{a}_{\tilde{2}})Z_2(\tilde{a}_{\tilde{1}},\tilde{b}_{\tilde{1}})^3J(a_1)J(a_2)J(a_3)\tilde{J}(\tilde{a}_{\tilde{1}},\tilde{b}_{\tilde{1}})\Big).
\label{eq:solid torus rel.}
\end{align}
 By using the solid torus equations \eqref{eq:solid torus rel.}, we can compute a couple of expectation values.
The first example is the expectation value of a duality defect on $S^2\times S^1$ as shown in Figure \ref{fig:S2timesS1}. 
We can cut out one solid torus from $S^2\times S^1$.
We apply the solid torus equation to this solid torus.
With the relations in  Figures \ref{fig:qdim} and \ref{fig:commutation_relation1}, we find the expectation value of the duality defect on $S^2\times S^1$ is one.
The second example is an expectation value of a $\Zb_2$ symmetry defect on a 2-dimensional disk whose boundary is a non-trivial $S^1$ cycle on the duality defect on $S^2\times S^1$ as shown in Figure \ref{fig:S2timesS1Z2}.
We use the solid torus equation to the duality defect configuration on $S^3$.
Then, we find that the expectation value is zero.

\begin{figure}[htbp]
  \centering
  \includegraphics[width=6cm]{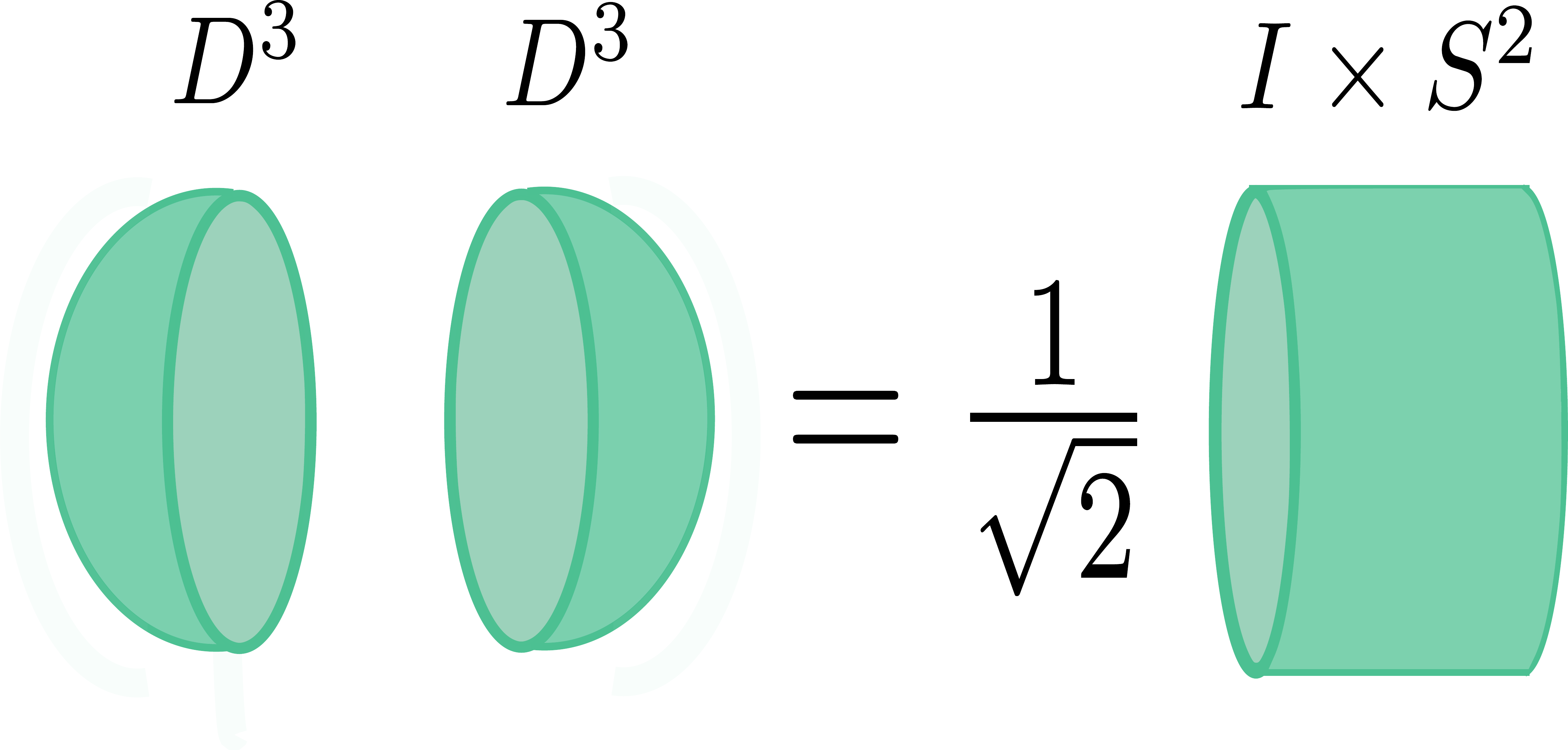}
  \caption{A schematic illustration of the crossing relation including two disconnected disks.}
  \label{fig:commutation_relation3_sch}
\end{figure}

\begin{figure}[htbp]
  \centering
  \includegraphics[width=13cm]{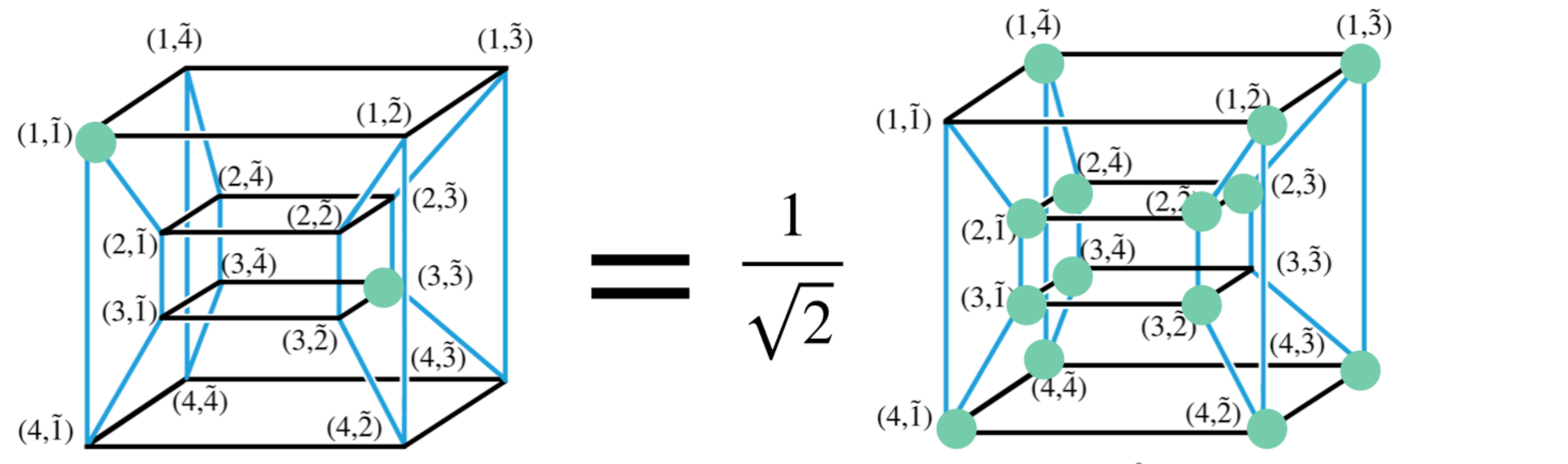}
  \caption{The configuration of Eq.~\eqref{commutationrelation3}.}
  \label{fig:commutation_relation3_place}
\end{figure}

We also find the crossing relation for duality defects placed on two disconnected 3-dimensional disks as depicted in Figure \ref{fig:commutation_relation3_sch}.
This relation is realized in a 16-cell as shown in Figure \ref{fig:commutation_relation3_place}.
This relation is expressed as
\begin{align}
	&s^8l^6W(a_1,a_2,a_3,a_4)D(a_1,\tilde{a}_{\tilde{1}})D(a_3,\tilde{a}_{\tilde{3}})\nonumber\\
	&=\frac{1}{\sqrt{2}}\sum_{\tilde{a}_2,\tilde{a}_4=0,1}s^8 l^8W(\tilde{a}_{\tilde{1}},\tilde{a}_{\tilde{2}},\tilde{a}_{\tilde{3}},\tilde{a}_{\tilde{4}},)D(a_1,\tilde{a}_{\tilde{2}})D(a_1,\tilde{a}_{\tilde{3}})D(a_1,\tilde{a}_{\tilde{4}})D(a_2,\tilde{a}_{\tilde{1}})D(a_2,\tilde{a}_{\tilde{2}})D(a_2,\tilde{a}_{\tilde{3}})\nonumber\\
	&\qq{}\qq{}\qq{}\times D(a_2,\tilde{a}_{\tilde{4}})D(a_3,\tilde{a}_{\tilde{1}})D(a_3,\tilde{a}_{\tilde{2}})D(a_3,\tilde{a}_{\tilde{4}})D(a_4,\tilde{a}_{\tilde{1}})D(a_4,\tilde{a}_{\tilde{2}})D(a_4,\tilde{a}_{\tilde{3}})D(a_4,\tilde{a}_{\tilde{4}}).\label{commutationrelation3}
\end{align}

\begin{figure}[htbp]
  \centering
  \includegraphics[width=8cm]{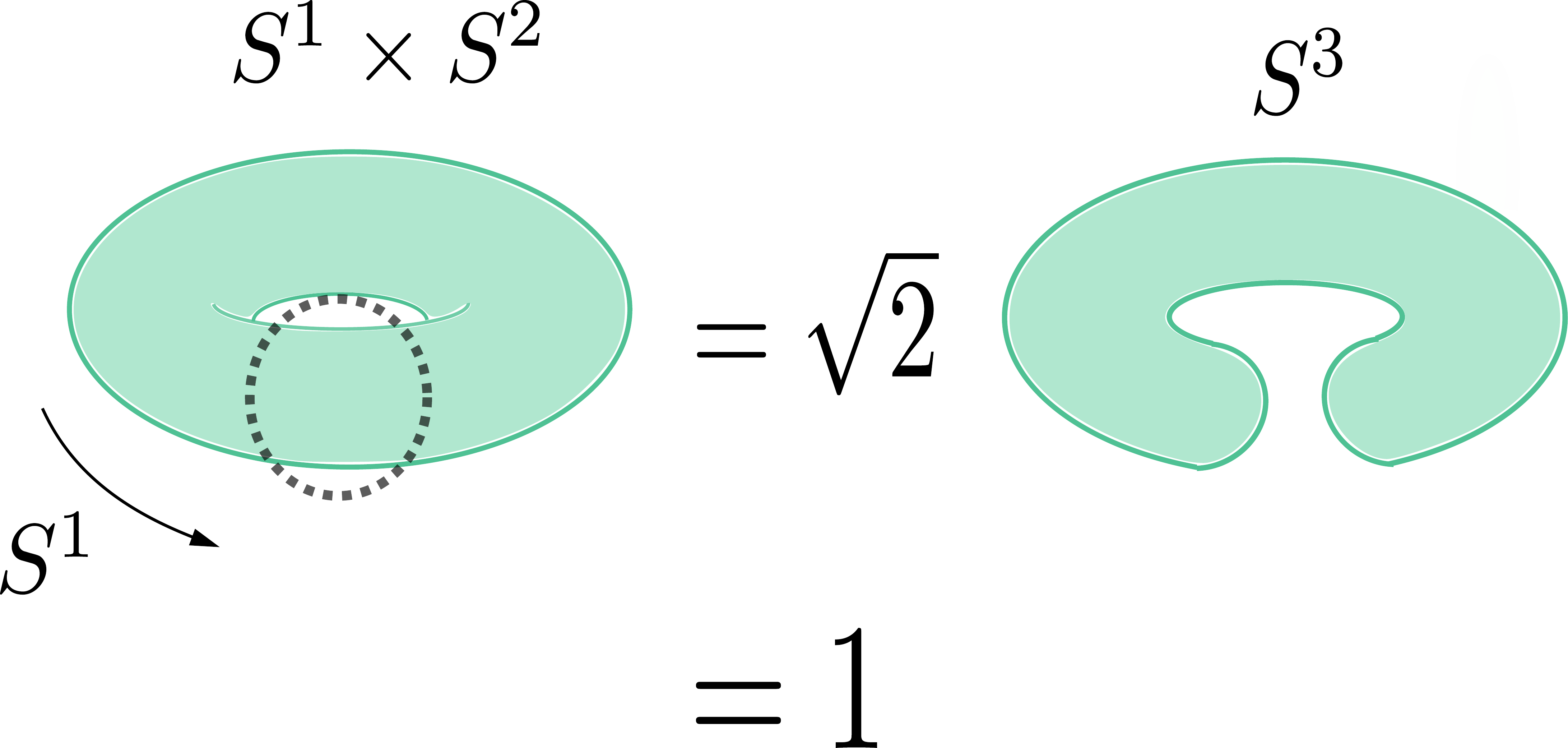}
  \caption{A calculation of an expectation value of a duality defect on $S^1\times S^2$. We use the commutation relation Figure \ref{fig:commutation_relation3_sch} to $S^1\times S^2$. Then, the $S^1\times S^2$ expectation value is equal to $\sqrt{2}$ times the $S^3$ expectation value. We see that the $S^1\times S^2$ expectation value is one by the relation Figure \ref{fig:qdim}.}
  \label{fig:S1timesS2}
\end{figure}

We can also calculate the expectation values of the duality defects by using this relation.
The first example is a duality defect on $S^3$.
We consider two disconnected duality defects on $S^3$.
We can relate this configuration and a $S^3$ duality defect by using Eq.~\eqref{commutationrelation3}.
Therefore, the expectation value of the duality defect on $S^3$ is $1/\sqrt{2}$.
This result is consistent with the result in Eq.~\eqref{eq;qdim}.
The second example is a duality defect on $S^1\times S^2$ as shown in Figure \ref{fig:S1timesS2}.
By using the crossing relation in Figure \ref{fig:commutation_relation3_sch} to $S^1\times  S^2$, we see that the $S^1\times S^2$ expectation value is one.

More generally one can calculate the expectation values of the duality defect on the connected sum by using the crossing relation in Figure \ref{fig:commutation_relation3_sch} as follows.  Let $\expval{X}$ the expectation values of the duality defects on a sub-manifold $X$.  Then the expectation value of the connected sum is given by
\begin{align}
  \expval{X\# Y}=\sqrt{2}\expval{X}\expval{Y}.
\end{align}
For example, $\expval{(S^1\times S^2)\#(S^1\times S^2)}=\sqrt{2}$.

\begin{figure}[htbp]
  \centering
  \includegraphics[width=6cm]{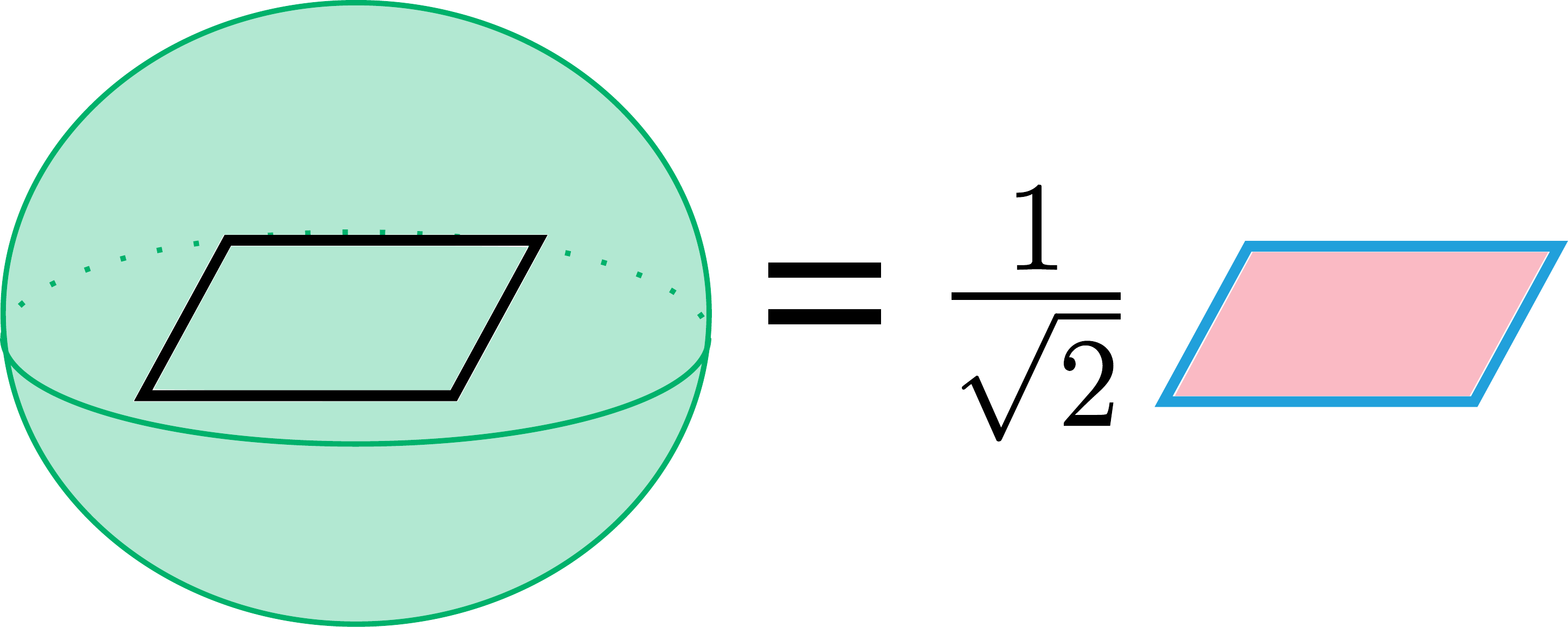}
  \caption{A schematic illustration of the action of a duality defect to a Wilson loop.}
  \label{fig:wilson comm1}
\end{figure}

\begin{figure}[htbp]
  \centering
  \includegraphics[width=7cm]{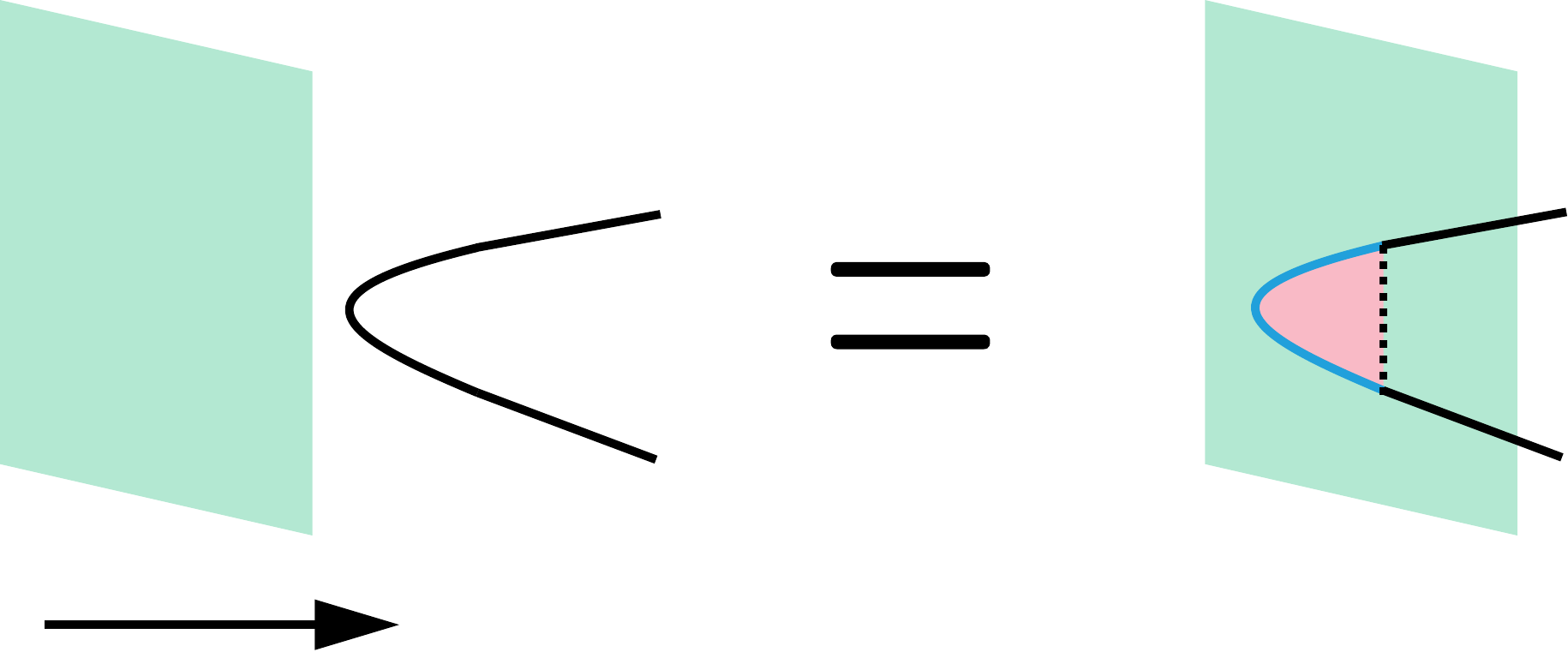}
  \caption{A schematic illustration of the general action of a duality defect to a Wilson loop.}
  \label{fig:wilson comm2}
\end{figure}
We explain the action of a duality defect to a Wilson loop.
When we place the duality defect around a plaquette Wilson loop, the relation in Figure \ref{fig:wilson comm1} is derived; a 't Hooft loop appears on the plaquette where the Wilson loop was located.
A 't Hooft loop by itself cannot be defined locally at the loop in this theory. It must be accompanied with a $\Zb_2$ symmetry defect on a surface whose boundary is the loop \cite{Gaiotto:2014kfa}.
The relation in Figure \ref{fig:wilson comm1} is expressed as
\begin{align}
	\sum_{\tilde{a}_{\tilde{1}},\tilde{a}_{\tilde{2}},\tilde{a}_{\tilde{3}},\tilde{a}_{\tilde{4}}=0,1}W(a_1,a_2,a_3,a_4)s^8l^8(-1)^{\tilde{a}_{\tilde{1}}+\tilde{a}_{\tilde{2}}+\tilde{a}_{\tilde{3}}+\tilde{a}_{\tilde{4}}}\prod_{(m,\tilde{n})\in U}D(a_{m},\tilde{a}_{\tilde{n}})&=\frac{1}{\sqrt{2}}W(1-a_1,a_2,a_3,a_4)s^4l^4.
\end{align}
In this equation, $U$ is defined as $U=\{1,2,3,4\}\times\{\tilde{1},\tilde{2},\tilde{3},\tilde{4}\}$.
The right-hand side $1/\sqrt{2}$ is the expectation value of the duality defect on $S^3$.
The sum of the $\Zb_2$ symmetry defect has already been evaluated. 
It is most likely that more general relations in Figure \ref{fig:wilson comm2} is satisfied.

\section{Conclusion and discussion}
\label{sec:conclusion}
In this paper, we construct topological defects in the 4-dimensional $\Zb_2$ lattice gauge theory, and discuss commutation relations between them.
We construct KWW duality defects and see these duality defects are non-invertible defects.  We also construct $\Zb_2$ symmetry defects and defect junctions.
We find several crossing relations among these defects.
The duality defects by themselves are not closed by these crossing relations, and thus $\Zb_2$ symmetry defects have to be introduced to close the crossing relations.
By using these crossing relations, we calculate the expectation values of a few defect configurations.

Here, we make some comments on the open questions and future issues.

When we impose the defect commutation relations and find the weights for a building block of duality defects in Sec.~\ref{sec:KWdefect}, 
we only consider configurations of defects where there is no ambiguous connection in which the duality defects are connected only by sites or links.
As for the deformations of duality defects including such ambiguous connections, we make the following interesting observations.
There are two classes of such deformations.
\begin{enumerate}
  \item There are relations similar to the defect commutation relation such as  Eq.~\eqref{eq;com relation}.
  \item There are relations similar to the solid torus crossing relation as in Figure \ref{fig:commutation_relation2_sch}.
\end{enumerate}
These two kinds of deformations of configurations are swapped when we replace active site/link and inactive site/link. 
A more detailed understanding of this phenomenon is a future problem.

In Sec.~\ref{sec:commutationrelations}, we compute the expectation values of $S^3$ and $S^1\times S^2$ duality defects. 
However, we do not compute those of other topologies such as $T^3$. 
These values are defined in this paper and computable in principle. Finding some nice way to compute these values and obtaining explicit values are future problems.  Since these expectation values are invariants of the embedding of a 3-dimensional closed oriented manifold into $\mathbb{R}^4$, this is also an important issue for mathematical implications.

In Sec.~\ref{sec:KWdefect}, we only consider the solution $D(a,\tilde{a})=(-1)^{a\tilde{a}}$ for simplicity, although we find that there are seven other solutions.
\begin{alignat}{2}
	D(a,\tilde a)=&-(-1)^{a\tilde a}
  ,\qquad
	&D(a,\tilde a)=&\pm(-1)^{(1-a){(1-\tilde a)}}\label{eq:duality_ans_2},\\
	D(a,\tilde a)=&\pm(-1)^{(1-a){\tilde a}}
  ,\qquad
	&D(a,\tilde a)=&\pm(-1)^{a{(1-\tilde a)}}\label{eq:duality_ans_4}.
\end{alignat}
The solutions \eqref{eq:duality_ans_2} satisfy $D(a,\tilde{a})=D(\tilde{a},a)$, while the solutions \eqref{eq:duality_ans_4} do not.
The same junction weights are obtained for the solutions \eqref{eq:duality_ans_2}, while the junction weights are different for the solutions \eqref{eq:duality_ans_4}:
\begin{align}
	J(a)&=-(-1)^a,\qquad
	\tilde{J}(b,c)=\sigma^x_{b,c}.
\end{align}
These solutions are completely equivalent to each other as far as one considers closed defect configurations in $\Rb^4$ since they are related by a gauge transformation in such defect configurations.
However, once one considers the space-time with non-trivial topology and defect configurations winding on the space-time, there may be some sign difference among the above solutions.  
Such a very slight difference may be useful to find a new characterization of topological phases.

Obtaining the operator formalism expression or the transfer matrix as described in \cite{Feiguin:2006ydp,Aasen:2016dop,Aasen:2020jwb} is an important future problem. 
Since there is the gauge symmetry in the lattice gauge theory, developing the operator formalism is not straightforward. It involves subtleties such as gauge fixing.

It has been found that a non-invertible symmetry exists at the critical point of the $\Zb_2$ gauge theory in \cite{Ji:2019jhk}.
This non-invertible symmetry is denoted by $Z_2^{\ (1)}\vee\widetilde{Z}_2^{\ (1)}$ in \cite{Ji:2019jhk}.
It is different from the non-invertible symmetry found in this paper since $Z_2^{\ (1)}\vee\widetilde{Z}_2^{\ (1)}$ in \cite{Ji:2019jhk} only includes codimension 2 defects while our non-invertible symmetry includes codimension 1 duality defects.

In this model, the phase transition is first order, and thus the continuum limit of this theory is a topological field theory.
Therefore, it is an important problem to find non-invertible topological defects such as duality defects in non-topological continuous quantum field theories.  
The structure of the non-invertible symmetry that we find in this paper should be quite universal. Thus we expect the same non-invertible symmetry appears in some continuum quantum field theories.
For example, $\mathcal{N}=4$ SU$(2)$ super Yang-Mills theory has the 1-form $\Zb_2$ center symmetry and duality similar to those in the pure $\Zb_2$ lattice gauge theory.
Therefore we expect that the $\Zb_2$ symmetry defects and the duality defects of $\mathcal{N}=4$ SU$(2)$ super Yang-Mills theory form the same non-invertible symmetry as one studied in this paper.

There are other interesting issues such as the analysis of QFT with non-invertible symmetry,
the mathematical structure that covers the crossing relations in Sec.~\ref{sec:commutationrelations}, and the anomaly of non-invertible symmetry.

\subsection*{Acknowledgement}
We would like to thank Wenjie Ji, Kantaro Ohmori, Yuji Tachikawa, and Seiji Terashima for useful discussions and comments.
The authors would also thank the Yukawa Institute for Theoretical Physics at Kyoto University.
Discussions during the YITP workshop YITP-W-21-04 on ``Strings and Fields 2021'' were useful to complete this work. 
The work of SY was supported in part by JSPS KAKENHI Grant Number 21K03574.

\bibliographystyle{utphys}
\bibliography{ref}
\end{document}